\RequirePackage{lineno}
\documentclass[a4paper,12pt]{article}

\usepackage[T1]{fontenc}
\usepackage[english]{babel}
\usepackage{epsfig,amsmath,subfigure,rotating}
\usepackage{longtable}
\usepackage{bm}
\usepackage{relsize}
\usepackage{graphicx}
\usepackage{multicol}
\usepackage{bicaption}
\usepackage{amsmath}
\usepackage{color}

\newcommand{\PANDA}{$\overline{\textrm{P}}\textrm{ANDA}$}
\newcommand{\bfp}{\boldsymbol{p}}

\newcommand{\genfit}{\textsc{Genfit2}}
\newcommand{\genfitone}{\textsc{Genfit}}

\begin{document}

\pagenumbering{arabic}

\pagestyle{empty}

\relax
\vskip 1.5 cm
\bigskip

\vskip 2.5 cm
\begin{center}
 {\huge \bf Implementation of \genfit~as an experiment independent track-fitting framework  }
  \end{center}
\vskip 1. cm
\medskip

\begin{flushleft}

      \large{Tadeas Bilka$^{[1]}$, Nils Braun$^{[2]}$, Thomas Hauth$^{[2]}$, Thomas Kuhr$^{[3]}$, Lia Lavezzi$^{[4,~5]}$, Felix Metzner$^{[2]}$, Stephan Paul$^{[6]}$, Elisabetta Prencipe$^{[7,~a]}$, Markus Prim$^{[2]}$,  Johannes Rauch$^{[6]}$, James Ritman$^{[7,~8]}$, Tobias Schl\"uter$^{[9]}$, Stefano Spataro$^{[5]}$}
\end{flushleft}
\vskip 1. cm
       \small{ [a] = corresponding author}\\
       \small{ [1] Charles University of Prague (CZ);[2]  Karlsruher Institut f\"ur Technologie (DE); [3] Ludwig-Maximilians-Univesit\"at M\"unchen (DE); [4] Institute of High Energy Physics (CN); [5] Universit\'a di Torino and INFN (IT); [6] Technische Universit\"at M\"unchen (DE); [7] Forschungszentrum J\"ulich (DE); [8] Ruhr-Universit\"at Bochum (DE); [9] LP-Research Inc.~(JP)}

{\vspace{0.5cm}}


%
\vskip 1.0 cm
\begin{abstract}
The  \genfitone~toolkit, initially developed at the Technische Universit\"at M\"unchen,  has been extended and modified to be more general and user-friendly. The new \genfitone, called \genfit, provides track representation, track-fitting algorithms and graphic visualization of tracks and detectors, and it can be used for any experiment that determines parameters of charged particle trajectories from spacial coordinate measurements. Based on general Kalman filter routines, it can perform extrapolations of track parameters and covariance matrices. It also provides interfaces to Millepede II for alignment purposes, and RAVE for the vertex finder. Results of an implementation of \genfit~ in basf2 and PandaRoot software frameworks are presented here.
\end{abstract}

\vskip 3 cm
\centerline{\large \today}
\cleardoublepage  
\setlength{\textheight}     {22.0  cm}
\setlength{\textwidth}      {15.5  cm}
\setlength{\baselineskip}   { 0.6  cm}
\setlength{\baselineskip}   { 20 pt}
\pagestyle{headings}

\tableofcontents
\section{Introduction}
\label{intro}

Track reconstruction is an essential part of most nuclear and particle physics experiments.

\genfit~\cite{gf2} is an experiment-independent modular framework for track-fitting and other related tasks. The project is open-source and available on GitHub\footnote{https://github.com/GenFit/GenFit.}.

All particle physics experiments need to identify and classify processes based on detector signals. Combining these signals to recover particle trajectories is the task called ``track reconstruction'': suitable collections of measurements must be combined into track candidates. Tracks must be fitted, and points of common origin or exit, vertices, must be found and fitted. Track finding and track fitting are not independent from each other. Steps of different levels of track finding
and refining alternate with track-fitting steps. Fitted tracks from different tracking subdetectors can be matched and combined into larger tracks; and single measurements can be appended to existing tracks, which might come from another subdetector. With provided suitable collections of measurements and detector geometry, all these tasks can be performed by \genfit.

\genfit~extends and improves the work previously performed by \genfitone~\cite{gf1}, the external tracking tool available in the PandaRoot releases since 2009. PandaRoot~\cite{pandaroot1, pandaroot2} is the official software framework of  the \PANDA~Collaboration~\cite{gsi}, which is building a future $\bar p p$ experiment investigating reactions where an antiproton beam with momentum between 1.5 GeV/$c$ and 15 GeV/$c$ impinges on a fixed target. \PANDA, which will be located at the international FAIR facility in Darmstadt (Germany), will consist of two main sections: the target spectrometer, which will have a solenoidal magnetic field ($\mathbf{B}$ = 2~T); and  a forward spectrometer, which will include a dipole field with 2 T$\cdot$m maximum bending power.  

The new \genfit~toolkit was already successfully used during a test of the Belle II high-level trigger architecture and the combined Belle II vertex detector readout architecture~\cite{bell2test}. It is used for recorded Phase 2 data and huge MC productions. Belle II~\cite{belle2} is a major upgrade of the Belle detector~\cite{belle}, which operates in a solenoid magnetic field of 1.5~T at an $e^+e^-$ collider in Tsukuba (Japan), running at a center of mass energy equal to that of the $\Upsilon(4S)$\footnote{The Belle II experiment concluded the Phase 2 data taking period on July 17$^{\rm th}$, 2018. During the Phase 3, planned  on spring 2019, the experiment will run at the  energy in the center of mass of the $\Upsilon(4S)$ and $\Upsilon(3S)$.}. During the aforementioned test it served for online track reconstruction in the data reduction stage of the high-level trigger, as well as for offline analysis. It was also used to supply input to the Millepede II alignment software~\cite{milleped}.

This paper documents the implemented classes, how to use them, and also presents performance results with the PandaRoot and the basf2 software frameworks. The latter is the official Analysis and Software Framework~\cite{basf2, basfkuhr} of the Belle II Collaboration.
This document is organized as follows: after this general introduction to \genfit, the motivation to build this toolkit is presented in the second Section. In Section 3 and 4, generalities about track parameterization, extrapolation and linearization are provided; the documentation of the \genfit~classes is given in Section 5, with the description of the general structure of this modular code. Section 6 describes the different track fitters available in \genfit~; Section 7 provided a short description of the interface between \genfit~and RAVE~\cite{reve}; Section 8 shows the performace in basf2 and PandaRoot;  finally, the work is summarized in the Conclusion.

\section{Motivation for an experiment-independent track-fitting toolkit~~~~}
\label{sec-repr}

In order to perform an analysis it is necessary to measure the  position of a particle with its uncertainty, and have a knowledge of the particle  momentum.

 A \emph{measurement} is a processed electronic signal recorded by a detector, which serves as input to the reconstruction algorithms.
Physics analyses generally work with particle \emph{tracks}. \emph{Tracks} can $e.g.$ be parametrized by an Euclidean position $\vec{x}$ and momentum $\vec{p}$ with the corresponding covariance matrix.  To form a track from a set of measurements, two tasks have to be performed: \emph{track-finding} and \emph{track-fitting}.  \emph{Track-finding} groups measurements which are likely to originate from the same particle into track candidates. This is a task on its own, and will not be detailed here. \emph{Track-fitting} takes the track candidates as input, and computes track parameters at any given point along the track, together with the covariance matrix of the track parameters, taking in consideration effects due to particle interactions with materials and the magnetic fields.

Several high-energy-physics experiments implement their own track fitters; however, they use similar algorithms, such as the well known \emph{Kalman filter}~\cite{KF}. The idea of \genfit~is to provide an open-source, modular and extensible framework which is capable of performing track-fitting and other related tasks, and can easily be adapted to various experimental setups. Smaller experiments, which do not have the manpower to develop their own track fitter, or new experiments which need a working tool to do research and development, are encouraged to use \genfit.

\section{Track parameterization and extrapolation}
\label{tobi:sec}
In the absence of interaction, the trajectory of a particle with charge
$q$ in a magnetic field $\mathbf{B}$ is determined by the equations of
motion, given its momentum vector $\mathbf{p}$, in a single point along
the track.  In order to describe a track, its trajectory has to be
parameterized.  In a general geometry, and inside a magnetic field, it is
most useful to use the distance along the track $s$ as free parameter,
and then give the values:
\begin{equation}
  \label{eq:1}
        (q/p,u,v,u',v')
\end{equation}
where $u,v$ are the rectangular coordinates of a plane which the
track intersects at distance $s$, and $u',v'$ are the direction cosines,
$i.e.$ the projections of the direction of momentum on the coordinate
axes.  If the plane is that of a planar detector, it is most useful to
have $u$ (and $v$) coincide with the coordinate(s) measured by the detector.

\genfit~extrapolates tracks using the standard Runge-Kutta-Nystr\"om
method~\cite{abramowitz+stegun} as modified by
Bugge and Myrheim to carry along the Jacobian
matrix~\cite{Myrheim:1979ng,Bugge:1980iv}.  The code shares the same
pedigree as the STEP extrapolation code used in the ATLAS
experiment~\cite{Lund:2009zzc,Lund:2009zzd}. The corresponding class in \genfit ~is {\tt RKTrackRep}.

During the Runge-Kutta extrapolation, global track parameters are
used, namely $q/p$, the spatial coordinates $\boldsymbol{x}=(x,y,z)^T$
and the unit tangent vector pointing along the track, $i.e.$ the
derivative of $\boldsymbol{x}$ with respect to the track length $s$.

In order to reduce bias of the final fit result due to inaccurate seed values, which can lead to failed linearization of the Kalman filter, usually several fitting steps are done until the fit converges. For example, we can assume here that the track is $e.g.$ processed three times in forward and backward directions,  giving an accurate estimation of the track parameters at the starting position of the track. The last backward update of one iteration is usually taken as a starting value for the next iteration.
In order to minimize the bias of the starting values, their covariance has to be very large. However, if it is too large, numerical problems can arise. It is practical to multiply the covariance of the starting value by a factor of 500 up to 1000 between 2 iterations. This factor can of course be adjusted by the user, depending on the specific case.

The fitter has a minimum and maximum number of iterations, which are assigned default values of 2 and 4, respectively, or the user can assign other values.  As soon as the minimum number of iterations has been reached, the tool checks if the $p$-value has changed by less than a certain amount with respect to the previous iteration. The default value to check if the fit should perform a next iteration is set to 10$^{-3}$. However, tracks with a $p$-value close to zero are often considered as converged with this criterion, while the $\chi^2$, albeit big, is still changing significantly, indicating that the fit is still improving. This situation occurs often for tracks which are given bad seed values. Thus, a non-convergence criterion has been introduced: if the relative change in  $\chi^2$ from one iteration to the next is more than 20\%, the fit will continue. The value of this non-convergence criterion can be defined by the user.

After the extrapolation has reached the target surface, global coordinates are converted to local coordinates (see Eq.~\eqref{eq:1}), and the Jacobian for the extrapolation is calculated from the Jacobian matrices of the extrapolation steps $J_i$ together with those of the coordinate changes local$\to$global on the starting plane $A$, and global$\to$local on the target plane $B$ as:
\begin{equation}
  \label{eq:extrap-jacobian}
  F_{A\to B}=J^{B}_{\textrm{global}\to\textrm{local}}J_nJ_{n-1}\cdots
  J_1J^A_{\textrm{local}\to\textrm{global}}.
  \end{equation}
More details about the global$\to$local coordinate system transformation are given in the next Sec.~\ref{linearize} and Appendix~\ref{coordinates}.
\section{Linearization}
\label{linearize}
All implemented track fitting algorithms use a discrete model, where the particle travels from detector plane $A$ to detector plane $B$ such that the parameters on plane $B$ are given as a function of the parameters on plane $A$, $i.e.$ $\bfp_B = h_{A\to B}(\bfp_A)$, and the Jacobian of this transport is the derivative of this function. Following the notation in Ref.~\cite{fruhw2}, we use the symbol:
\begin{center}
$F_{A\to B} \equiv \left.\frac{\partial h_{A\to B}(\bfp)}{\partial
    \bfp}\right|_{\bfp=\bfp_A}$.
\end{center}
The covariance of the track parameters on plane $B$ can then expressed as:
\begin{equation}
  \label{eq:2}
  C_B=F_{A\to B} C_A F_{A\to B}^T + N_{A\to B},
\end{equation}
where $N_{A\to B}$ is the noise matrix evaluated by the track extrapolation algorithm.  The evaluation of $(\bfp_B,C_B)$ is referred to as prediction.

Each measurement $m_i$ gives rise to a residual $r_i$, the difference between the measurement expected based on the track parameters $\bfp_i$ and the measured value, which in general is given as:
\begin{equation}
  \label{eq:3}
  r_i = m_i - f_i(\bfp_i),
\end{equation}
where $f$ is the measurement function which calculates the expected measurement given a set of track parameters.  The measurement itself has a covariance matrix $V_i$.  Since our track parameterization uses planar coordinates, this can be brought into a simple linear form when the $u$ (and $v$ for two-dimensional detectors) direction and zero point of the plane - in which the track parameters are given - coincide with that of the measured coordinates,
\begin{equation}
  \label{eq:4}
  r_i = m_i - H_i \bfp_i.
\end{equation}
In this way, no further linearization is needed in order to apply the Kalman filter to this measurement model, and $H_i$ is a constant projection matrix for each detector.  In general, $H_i$ has to be replaced with the derivative of $f_i$.

The evaluation of $F_{A\to B}$ and $N_{A\to B}$ on the other hand cannot be eliminated by such a judicious choice of coordinates, and its value depends on the unknown track parameters.  Especially the first few hits pose a problem: the track parameters only become determined successively while the transport matrix already needs to be evaluated.

\genfit~has two approaches to this problem:
\begin{itemize}
\item use the current result of the track fit, which may lead to imprecise or false results   (it is the {\texttt{SimpleKalmanFitter}} described in Sec.~\ref{fitters});
\item use information from the track finders to estimate an initial
  set of transport and noise matrices, then after each iteration of
  the fit prepare a new set of matrices which is used in the next
  iteration (it is the \texttt{ReferenceKalmanFitter} described in  Sec.~\ref{fitters}).
\end{itemize}
\section{Structure of the code}
\label{cap1}

\genfit~is a C++ object-oriented modular tool, structured as shown in Fig.~\ref{fig1}. It provides general track-fitting tools, which are based on three pillars:

\begin{figure*}[ht] 
  \begin{center}
    \includegraphics[width=42pc]{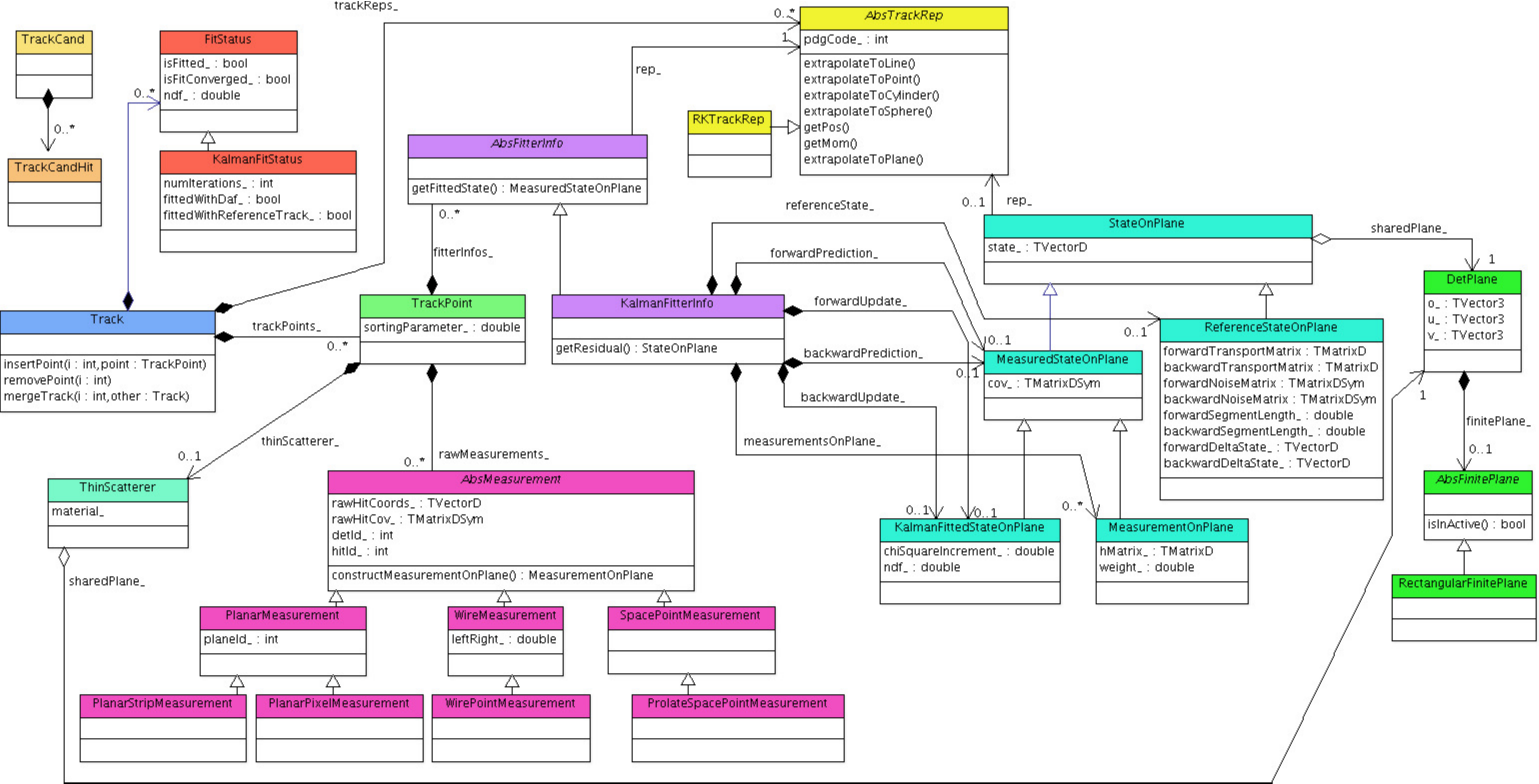}
    \caption{\label{fig1}  Structure of the \genfit~package~\cite{gf2}.}
  \end{center}
\end{figure*}
\clearpage
\begin{itemize}
\item measurements; 
\item track representations;
\item fitting algorithms. 
\end{itemize}

$Measurements$ serve as objects containing measured coordinates from a detector. They provide functions to construct a (virtual) detector plane and to provide measurement coordinates and covariance in that plane.   The abstract base class {\tt AbsMeasurement} defines the interface. A measurement can be one-dimensional ($e.g.$ a line segment measured by a semiconductor strip detector), two-dimensional ($e.g.$ a position on a semiconductor pixel detector or a drift isochrone of a wire detector), or three-dimensional ($e.g.$ a point in space measured by a Time Projector Chamber (TPC)). \genfit~comes with predefined measurement classes for various detector types, which are  called \texttt{PlanarMeasurement}, \texttt{WireMeasurement} and \texttt{SpacePointMeasurement}.
\begin{figure}
    \centering
    \includegraphics[width=0.5\textwidth]{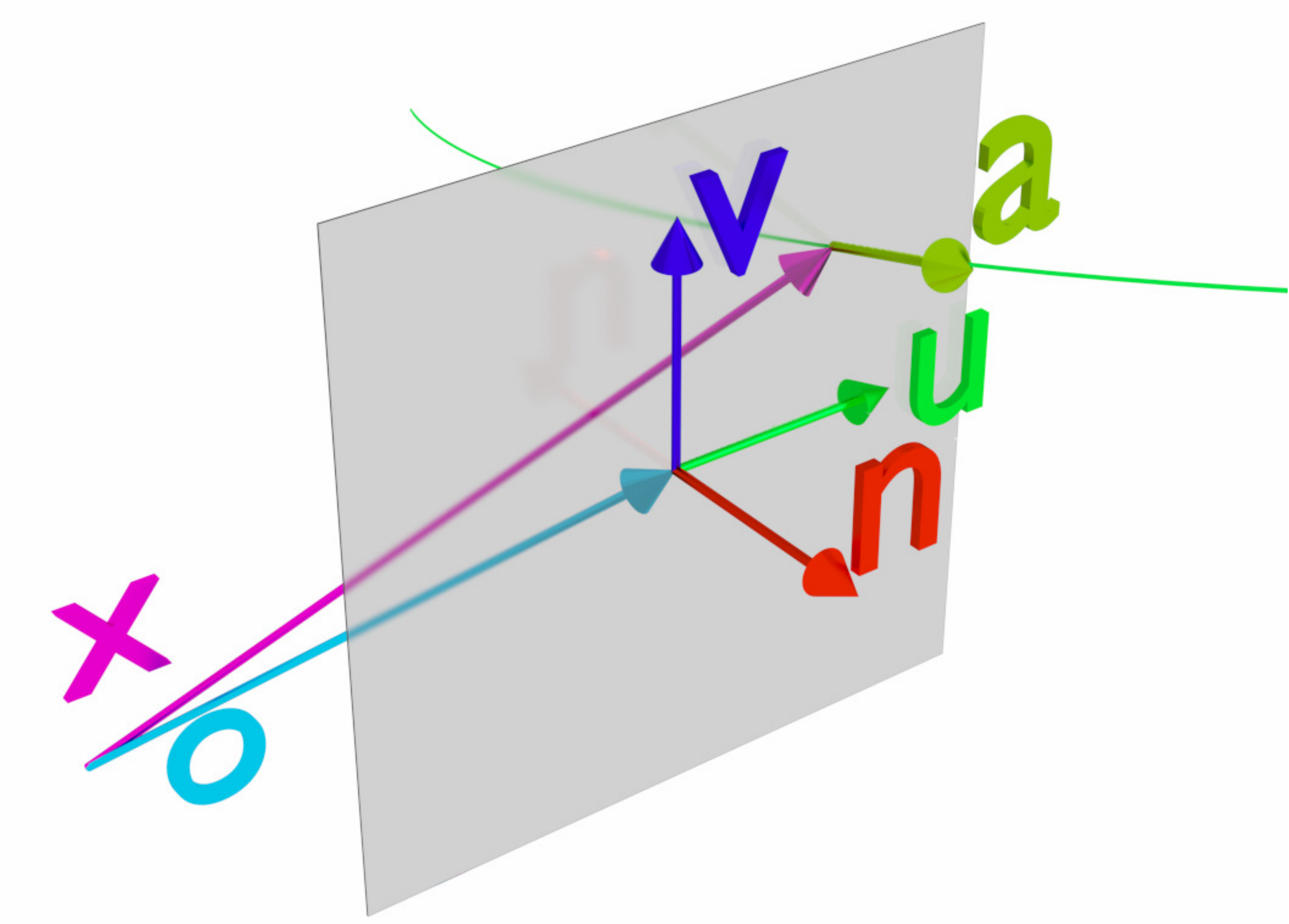}
    \caption{Track parametrization in the \genfit~class {\tt RKTrackRep}.}
    \label{fig:detPlane}
  \end{figure}		

In \genfit, state vectors ${\bf p}$ and measurements $m$ are given in local coordinate systems (see formulae in Appendix A), which are defined by detector planes and implemented in the {\tt DetPlane} class. A detector plane has an origin $\vec{o}$, and is spanned by two orthogonal unit vectors $\vec{u}$ and $\vec{v}$. The normal vector $\vec{n}$ is then equal to $\vec{u} \times \vec{v}$ (see Fig.~\ref{fig:detPlane}). Measurements provide a matrix that can project the state ${\bf p}$ and covariance $C$ into the measurement coordinate system of $m$ and its covariance $V$. 
For the Kalman filter mathematics, which are detailed in Appendix~A, the  state ${\bf p}$ and covariance $C$ have to be projected into the measurement coordinate system of $m$ and $V$.  These projections are described by matrices, as in Eqs.~\ref{hmatrix1}-\ref{hmatrix3}.

If we consider a silicon strip detector with strips parallel to $\vec{u}$, the coordinate $v$ in the local plane system reads as:
\begin{align}\label{hmatrix1}
 m = \left(v\right).
\end{align}
The $H$ matrix now simply has to project out the $v$ component of $p$:
\begin{align}\label{hmatrix2}
 H = \left(\begin{matrix} 0 & 0 & 0 & 0 & 1 \end{matrix} \right).
\end{align}
Now, $Hp$  is the projected state, and $HCH^T$ is the projected covariance.

For a detector which measures $u$ and $v$, the following applies:
\begin{align}\label{hmatrix3}
 m &= \left(\begin{matrix} u & v \end{matrix} \right) \\
 H &= \left( \begin{matrix}
             0 & 0 & 0 & 1 & 0 \\
	     0 & 0 & 0 & 0 & 1
            \end{matrix}
      \right).
\end{align}
For planar detectors, the detector plane is given by the detector geometry, whereas for {\tt wire} and {\tt spacepoint} measurements, so-called virtual detector planes are constructed by extrapolating the track to the point of closest approach (POCA\footnote{POCA  stands for Point of Closest Approach.}) to the spacepoint or wire. Additional detectors effects can be used here, allowing one to compensate for detector deformations like plane bending, wire sagitta, and misalignment during this procedure.

Drift-time corrections and cluster fits (track-dependent clustering) are possible.

\subsection{Track data structure}
\label{trkdata}
All per-track data is kept in the $Track$ object. It holds a set of {\tt TrackPoint} objects, which can contain $Measurements$, $FitterInfo$ objects, which hold all fitter-specific information, and thin scatterers (currently only used in the GBL fitter, see Section~\ref{gblsec}).

A $Track$ contains one or more track representations, representing the particle hypotheses that should be used for the fit. One of them must be selected as the cardinal track representation. This can be done by the user or by \genfit, which selects the track representation that best fits the measurements ($e.g.$ it has the lowest $\chi^2$). The $Track$ also contains a $FitStatus$ object, which stores general information (number of iterations, convergence, etc.) and fit properties ($\chi^2$, NDF, $p$-value, track length, etc.).
The track candidate ({\tt TrackCand}) serves as a helper class, basically storing indices of raw detector hits in $TrackCandHit$ objects, which can also be overloaded by the user to store additional information. $WireTrackCandHit$ objects can store the left-right ambiguity.

A possible use case for this possibility to incorporate within these objects is the pattern recognition: algorithms can work on raw objects that have a lighter weight content than a $Measurement$, or implement other features needed by the pattern recognition, but not necessarily by \genfit. 

\subsection{Pruning}
\label{sec-pruning}
The \genfit~ class {\tt MeasurementFactory} can build a $Track$ from a {\tt TrackCand}. This $Track$ can then be processed by the various fitting algorithms. After fitting,  $Track$ objects contain a lot of data: track representations, {\tt TrackPoint} objects with measurements, $FitterInfo$ objects, etc. Usually, not all of this information needs to be stored on disk. For example, for physics analysis only parameters related to the vertex and information on the track quality are necessary.
The user can decide which data to keep, $e.g.$ only the fitted state of the first {\tt TrackPoint}, and only for the cardinal track representation.

\subsection{Handling of multiple measurements in one TrackPoint}
Wire detectors measure a drift isochrone (a cylinder, in the simplest case). Intersected with the detector plane, this  gives two (one dimensional) measurements, since the particle could have passed on either side of the wire.
\genfit~ provides the possibility to store several measurements of the same type in one  {\tt TrackPoint}\footnote{This feature allows the DAF to assign weights to the {\tt TrackPoint}. More information related to the DAF are in Sec.~\ref{daf}.}. These tracks can also be fitted with the Kalman filter, therefore \genfit~provides several options how to handle multiple measurements:
\begin{itemize}
 \item $average$: the average of the individual measurements is calculated. This option is primarily used for the {\tt DAF} (see Sections~\ref{tobi:kal} and \ref{daf}).
  \item $closest~ to~ prediction$: the measurement which is closest to the state prediction is selected.
  \item $closest~ to~ reference$: the measurement which is closest to the reference track is selected. This can only be used with the Kalman filter with reference track.
  \item $closest~ to~ prediction/reference~   for~ wire~ measurements$: if the {\tt TrackPoint} has one $WireMeasurement$, select the side which is closest to the prediction/reference. Otherwise, it makes use of the $average$.
\end{itemize}

For wire measurements, it turned out that selecting the side which is closest to the state predictions is often the best option.
The average weigthed measurement $m$ and its covariance $V$ are calculated from the individual measurements $i$ as follows:
\begin{align}
  \begin{split}
    V & = \left( \sum_i \left(  w_i V_i^{-1} \right) \right) ^{-1} \\
    m & = V \cdot  \sum_i \left( w_i V_i^{-1} m_i \right) \\
  \end{split}
  \label{eq-averageMeasurement}
\end{align}

\subsection{Runge-Kutta track representation}
The Runge-Kutta track representation ({\tt RKTrackRep}) is based on a Runge-Kutta extrapolator adapted from Geant3~\cite{g3rep}. An abstract interface class interacts with the detector geometry. Implementations using the Root class {\tt TGeoManager}~\cite{g4rep_1} and Geant4 class {\tt G4Navigator}~\cite{g4rep_2} are available.

The state ${\bf p}$ is parametrized with 5 coordinates in a local plane coordinate system, as already shown in Fig.~\ref{fig:detPlane}. The cartesian position $\vec{x}$ and direction $\vec{a}$ translate into plane coordinates according to the following equations:
  \begin{align}
    \begin{split}
      {\bf p}        &= \left( q/p, u^\prime, v^\prime, u, v \right)^T \\
      u^\prime &= \frac{\vec{a} \cdot \vec{u}}{\vec{a} \cdot \vec{n}} \\
      v^\prime &= \frac{\vec{a} \cdot \vec{v}}{\vec{a} \cdot \vec{n}} \\
      u  &= \left( \vec{x} - \vec{o} \right) \cdot  \vec{u} \\
      v  &= \left( \vec{x} - \vec{o} \right) \cdot \vec{v}. \\
    \end{split}
    \label{eq:RKParametrization}
  \end{align}
where $q$ is the charge of the track, and $\mathbf{p}$ its momentum. During fitting, material properties are used to calculate the following effects: energy loss and energy-loss straggling for charged particles according to the Bethe Bloch formula (the code is based on the code ported from Geant3); multiple scattering (according to Ref.~\cite{ref7} or using the Highland-Lynch-Dahl formula~\cite{HLD}), where the full noise matrix is calculated; and soft Bremsstrahlung energy loss and energy-loss straggling for $e^-$ and $e^+$ (the code is based on the code ported from Geant3).
The step sizes used for the Runge-Kutta extrapolation should be as large as possible to avoid unnecessary computation, while still being small enough to keep errors reasonably small. An adaptive step-size calculation is done in the {\tt RKTrackRep}, taking field inhomogeneities and curvature into account. To calculate material effects correctly, extrapolation stops at material boundaries, and the step size is limited so that a maximal relative momentum loss in the material is not exceeded. {\tt RKTrackRep} provides different methods to find the POCA  of the track to non-planar measurements. These are used to construct virtual detector planes:
\begin{itemize}
\item  $extrapolate~to~point$ finds the POCA of the track to a given spacepoint. The virtual detector plane contains the spacepoint and the POCA, and it is perpendicular to the track direction. A weight matrix can be used as a metric, defining the space in which the POCA will be calculated. By default, the inverted 3D covariance of a spacepoint measurement is used as a metric, which gives correct fitting results also for spacepoints with arbitrary covariance shapes. Again, the virtual detector plane contains the spacepoint and the POCA, and it is perpendicular to the track direction in the space defined by the metric;

\item  $extrapolate~to~line$ finds the POCA of the track to a given line or wire. The virtual detector plane contains the line and the POCA. This routine is used for fitting wire measurements. 
\end{itemize}

The intersection of the virtual detector plane and the measurement covariance gives the covariance in the plane.

{\tt TrackRep} combines track parameterization and track extrapolation code.
The state can be extrapolated through material and the magnetic field, to:
\begin{itemize}
\item  detector planes;
\item the POCA to points, lines, cylinders, and cones;
\item or by a certain distance.
\end{itemize}
The state and covariance can be converted to/from a state defined in the global cartesian coordinate system, $(\vec{\bf x}~~  \vec{\bf p})$, as detailed in Appendix A. The abstract base class {\tt AbsTrackRep} defines the interface, and \genfit~ ships with the {\tt RKTrackRep} as a concrete implementation.
\section{Fitter implementation}
\label{fitters}
 Track representations combine track-parameterization and track-extrapolation code. Fitting algorithms use the measurements and track representations to calculate fit results, which are stored in corresponding objects in the $Track$ objects.

 Four different track-fitting algorithms are currently implemented in \genfit: 
\begin{itemize}
\item two smoothing Kalman filters, one which linearizes the transport around the state predictions, and one which linearizes around a reference track; 
\item a deterministic annealing filter ({\tt DAF}); 
\item a general broken lines (GBL) fitter.
\end{itemize}

\genfit~provides the possibility to store several measurements of the same type in one {\tt TrackPoint}, mainly for using the DAF to assign weights to them. Wire measurements also produce two {\tt MeasurementOnPlane} objects, representing the passage of the particle on either side of the wire. These tracks can also be fitted with the Kalman filters. 
\subsection{Kalman filter}
\label{tobi:kal}
The Kalman filter is a progressive algorithm that successively adds the
information from each detector hit to the estimated track parameters
using a linear model of measurement and track propagation.
It uses a series of measurements observed over time, containing ``noise'' (random variables) and other inaccuracies, and produced estimates of unknown variables; then iterates a least square method in 3 steps: prediction, update, and smoothing.

Smoothing~\cite{fruhw3} is a standard technique: the track is fitted in forward and backward direction, where predictions and updates are saved in the $FitterInfos$ at each {\tt TrackPoint}. With both of these, smoothed track states can be calculated. The weighted average between forward and backward prediction gives the unbiased state, whereas the average between prediction in one direction and update of the opposite direction results in the biased smoothed state.
This smoothed track gives more accurate states than either forward or backward updates alone.

The Kalman filter method introduced in \genfit~is
widely documented. We refer specifically to Ref.~\cite{fruhw2},
and here we only document the specifics of our implementation which deviates
in two ways.  First, all matrix operations are performed in a way
which explicitly maintains the symmetry of the covariance matrices at
each step in the calculation, thus minimizing the number of operations
and avoiding inconsistencies due to round-off errors.  Second, an
alternative implementation is available which implements a square-root
filter which additionally guarantees that the covariance matrices
remain positive-definite while improving the robustness against
round-off errors or vastly different matrix entries which can
especially happen at the beggining of the track fit, where individual
track parameters only become known successively.

The Kalman filter performs a succession of fits (called ``updates''),
moving from the innermost hit along the series of hits
identified by the track finders.  Then the result of this first
iteration of fits in the outwards direction is used as the starting
value for the next iteration in which the fit proceeds
in the backwards direction.  In
order to avoid biasing the inwards fit, the covariance matrix of the
result of the outwards fit is multiplied by a large factor and the
off-diagonal entries are discarded.  At each {\tt TrackPoint} the
results from inward and outward fit are combined by averaging the
updated state of the outward fit and the prediction of the inward fit.
This averaging is called smoothing, but it should be kept in mind that
also for the smoothed track, the track parameters change discretely at
each measurement.  Averaging is thus an important operation in
\genfit, and some care has gone into a numerically stable, efficient
implementation.  This implementation is detailed in
Appendix~\ref{app:averaging}.

Given the track parameters $\bfp_{i-1|i-1}, C_{i-1|i-1}$ at {\tt
  TrackPoint} $i-1$ after the preceding update step, we evaluate the
prediction at plane $i$:
\begin{align}
  \bfp_{i|i-1} & = h_{(i-1)\to i}(\bfp_{i-1|i-1}),\\
  C_{i|i-1} & = F_{(i-1)\to i}C_{i-1|i-1}F_{(i-1)\to i}^T + N_{(i-1)\to i},
\end{align}
and the residual of the prediction:
\begin{equation}
  \label{eq:14}
  r_{i|i-1} = m_i - H_i\bfp_{i|i-1}.
\end{equation}
Taking care to maintain the symmetry properties of the matrices, we
can then write the updated state and covariance as:
\begin{align}
  \label{eq:15}
  \bfp_{i|i} & = \bfp_{i|i-1} +
                   C_{i|i-1}H_i^t(V_i+H_iC_{i|i-1}H_i^T)^{-1}
                   r_{i|i-1},\\
  C_{i|i} & = C_{i|i-1} - C_{i|i-1}H_i^T(V_i+H_iC_{i|i-1}H_i^T)^{-1}H_iC_{i|i-1}.
\end{align}
This summarizes the formalism of the Kalman filter.  The only
difference between the so-called \texttt{SimpleKalmanFitter} and the
\texttt{ReferenceKalmanFitter} lies in the evaluation of the transport
and noise matrices, as explained above.  Nevertheless, we have
implemented an alternative equivalent formalism, using square-roots of
covariance matrices to further improve numerical accuracy and
resilience.  This implementation is described in
Appendix~\ref{sec:proof-square-root}.  Employing it allows significant
increases in the blow up factor applied to the covariance matrix on
fitter returns.

During the first iteration, the reference track is constructed by extrapolating the seed state from  measurement to measurement. In the following iterations, the smoothed states of the previous iterations are taken as a new reference states.
A $\chi^2$ value is calculated according to:
    \begin{align}
      \begin{split}
	r_{k} & = {\bf p}_{k|n} - {\bf p}_{k,r} \\
	\chi^2_{k} & = \sum_i \frac{(r_{k|n, i})^2}{V_{k|n, i,i}}. \\
      \end{split}
      \label{eq-chi2}
    \end{align}
    If the $\chi^2$  is below a certain value (default assigned value is 1), the corresponding reference state will not be updated to save computing time.

The same convergence criteria as for the standard Kalman filter are used.
Additionally, the fit converges if the reference states do not change anymore, according to Eq.~[\ref{eq-chi2}].

The choice of the linearization point can significantly affect the performance of the track fit. In particular, if  the first few hits lead to a large misestimate of the curvature, state predictions may stay very far from the actual trajectories. Consequently, linearizing around them is not optimal; material and magnetic field lookups are also not done at the proper places. It is therefore common to linearize around reference states~\cite{fruhw}, which are calculated by extrapolating the start parameters to all {\tt TrackPoint} objects. At later iterations, the smoothed states from the previous iterations are used as linearization points. However, if the change would be very small, reference states are not updated, saving computing time. It is also possible to let the fitter sort the measurements along the reference track, which can improve the fitting accuracy. In addition to the convergence criteria detailed above, the fit is regarded as converged if none of the reference states has been updated since the previous iteration.
\subsection{Deterministic annealing filter}
\label{daf}
The deterministic annealing fitter (DAF)~\cite{fruhw2} is a powerful tool for the rejection of outlying measurements. It is a Kalman filter that uses a weighting procedure between iterations based on the measurement residuals to determine the proper weights. The user can select which of \genfit~two Kalman filter implementations should be used, and specify the annealing scheme.

The DAF is also perfectly suited to resolving the left-right ambiguities of wire measurements. However, a problem can occur: the weights of the {\tt MeasurementOnPlane} objects must be initialized. The basic solution is to initialize both left and right measurements with a weight of 0.5. Effectively, the wire positions are taken as measurements in the first iteration, and their covariance is twice the mean of the individual covariances. So, all the wire positions have the same covariance, no matter how far from the actual trajectory they are. This systematically false estimate of the covariances biases the fit. \genfit~implements a novel technique to initialize the weights that improves the fitting efficiency: measurements with larger drift radii are assigned smaller weights, leading to larger covariances since the wire position is expected to be farther away from the trajectory. The mathematical espression used in the code is:
    \begin{align}
      \begin{split}
	w & = \frac{1}{2} \left( 1 - \frac{r_{\text{drift}}}{r_{\text{drift, max}}}  \right) ^2, \\
      \end{split}
      \label{eq-weightInit}
    \end{align}

    where $w$ is the given weight, and $r_{drift}$ is the distance of the measurement from trajectory.

    In contrast, measurements with smaller drift radii, which are closer to the trajectory, get larger weights. When the annealing scheme has been passed, convergence is checked: if the absolute change of all weights is less than $\rm{10^{-3}}$ (user configurable), the fit is regarded as converged. Otherwise another iteration is done, until the fit is converged or a maximum number of iterations is reached.

\subsection{Generalized broken lines fitter}
\label{gblsec}
The generalized broken lines (GBL) method of track fitting~\cite{gbl1, gbl2} was implemented especially for the purpose of alignment with Millipede II~\cite{millepede}. It is mathematically equivalent to the Kalman filter (with thin scatterers instead of continuous materials), and works with a large number of parameters, but fits tracks in their entirety in one step. For alignment purposes, \genfit~also provides a set of interfaces for alignment parameters and derivatives which can be implemented by the detector classes.

\section{Vertex reconstruction}

{\tt GFRave}, an interface to the vertex-fitting framework RAVE~\cite{reve}, has been implemented in \genfit. RAVE is a detector-independent toolkit for vertex reconstruction, originally developed for the CMS experiment~\cite{cms}. {\tt GFRave} takes full advantage of the \genfit~material model and the sophisticated algorithms of RAVE, allowing for precise and fast vertex reconstruction. GFRave is  part of the basf2 framework, where it is actively used. However, there is still no corresponding implementation in PandaRoot, which is using for the time being other vertex and kinematic fitters. 

\section{Performance}
\subsection{Application of the \genfit~standalone toolkit}
\label{perf}
Examples of what can be the result of a standalone \genfit~simulation are given in Figs.~\ref{picgf2}-\ref{fig:eventDisplay}, which display the screenshots of an event reconstructed with \genfit, performed by an algorithm based on {\tt Teve}: the reader can see that many features are available in the left-menu. Figures 5, 6, 7 show tests performed with a demonstration macro provided together with the \genfit~toolkit ($makeGeom.C$). By default, this test macro creates a geometry which comprises a 1 m$^3$ cube filled with air, and 4 layers of cylindrical silicon material with a radius of 1.0 cm, a length of 10 cm and a wall-thickness of 0.05 cm. The creation of $ measurements$ in the test macros is independent on the geometry, but the geometry is basically provided to evaluate material effects. One million events with a single pion in a solenoidal field of 2 T have been simulated. Different pion momenta have been simulated: 50 MeV/c, 100 MeV/c, 400 MeV/c, 1 GeV/c.   The information illustrated in the standalone \genfit~event visualization is related to the detector, hits, track with errors and track markers, the reference track, the result of the fit in forward and backward direction.

\subsection{Visualization of tracks in basf2}
\label{sec:visualisation}
\begin{figure*}[h!] 
\begin{center}
\mbox{
\includegraphics[width=32pc]{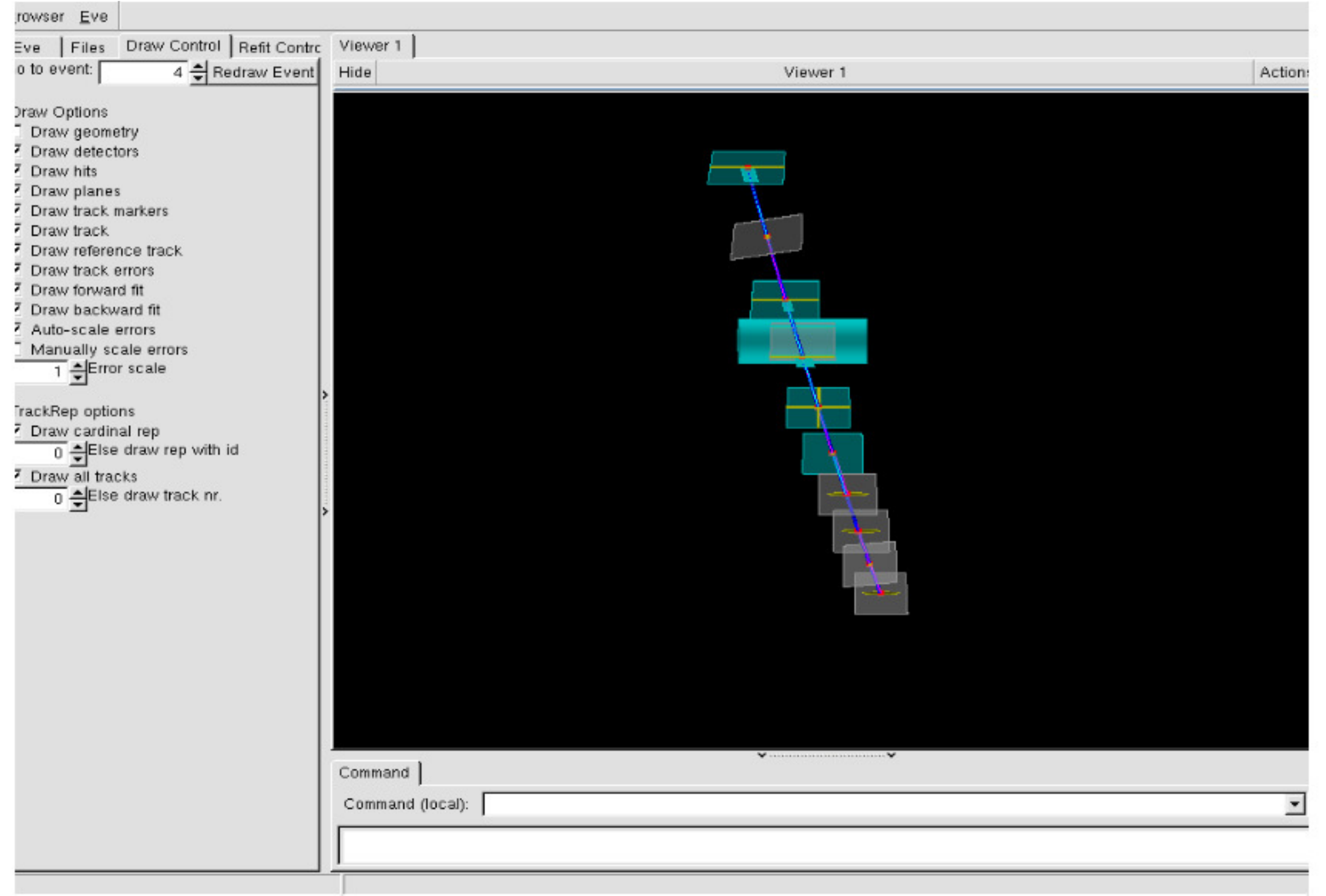}
}
\caption{\label{picgf2} Example of event visualization using the standalone \genfit~tool. Here a pion of 400 MeV/c is generated in a solenoidal magnetic field  $\mathbf{B}$= 2~T. }
\end{center}
\end{figure*}

\begin{figure}[ht]
        \centering
        \subfigure[Measurements with covariance (yellow), planar detectors and drift isochrones (cyan), respectively, and reference track (blue).] {\includegraphics[trim =  16cm 2.25cm 20cm 4.6cm, clip, width=0.4\textwidth]{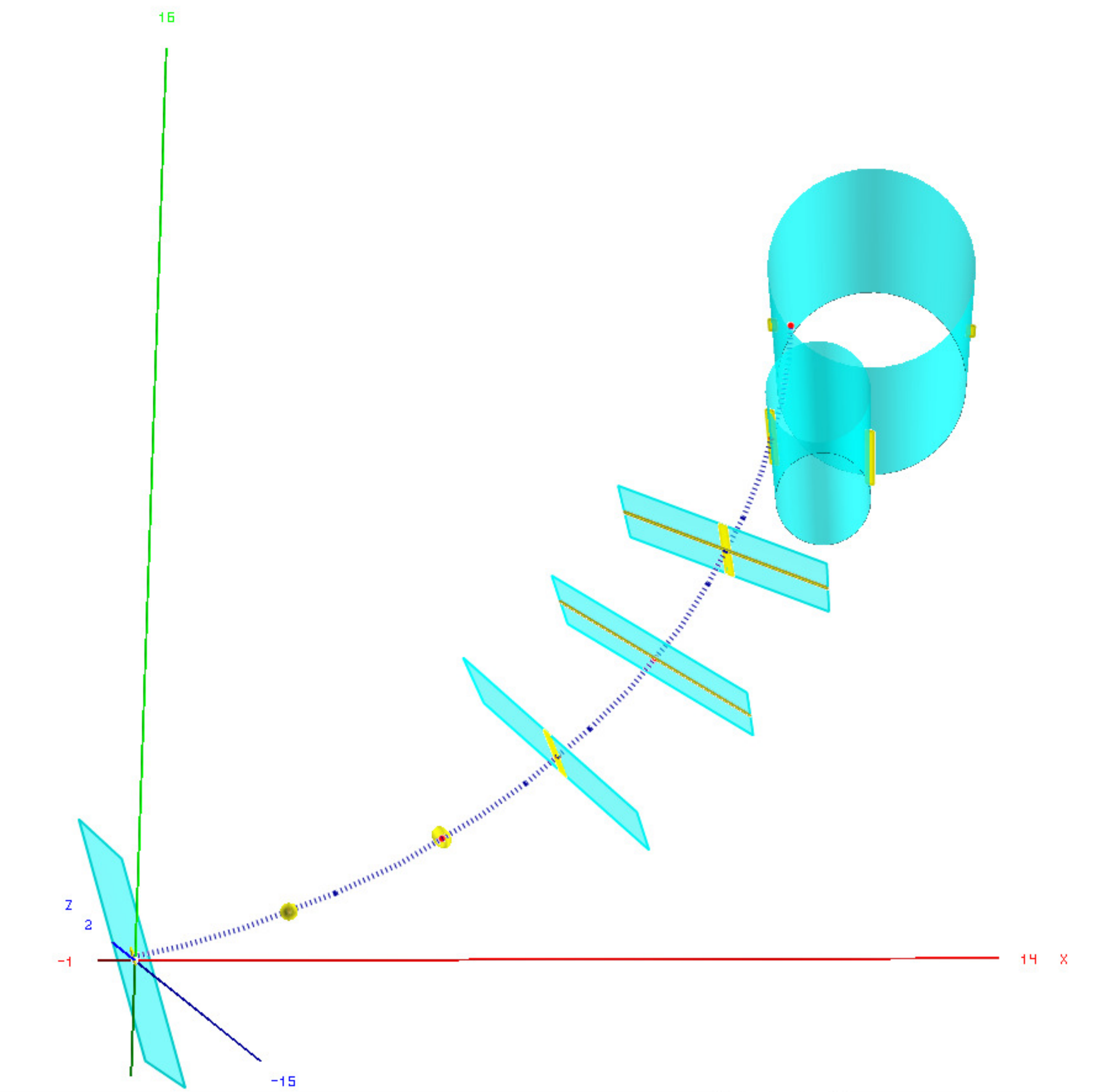}}\quad
        \subfigure[Detecor planes (grey). For the spacepoint- and wire-hits, virtual detector planes have been constructed.]{\includegraphics[trim =  16cm 2.25cm 20cm 4.6cm, clip, width=0.4\textwidth]{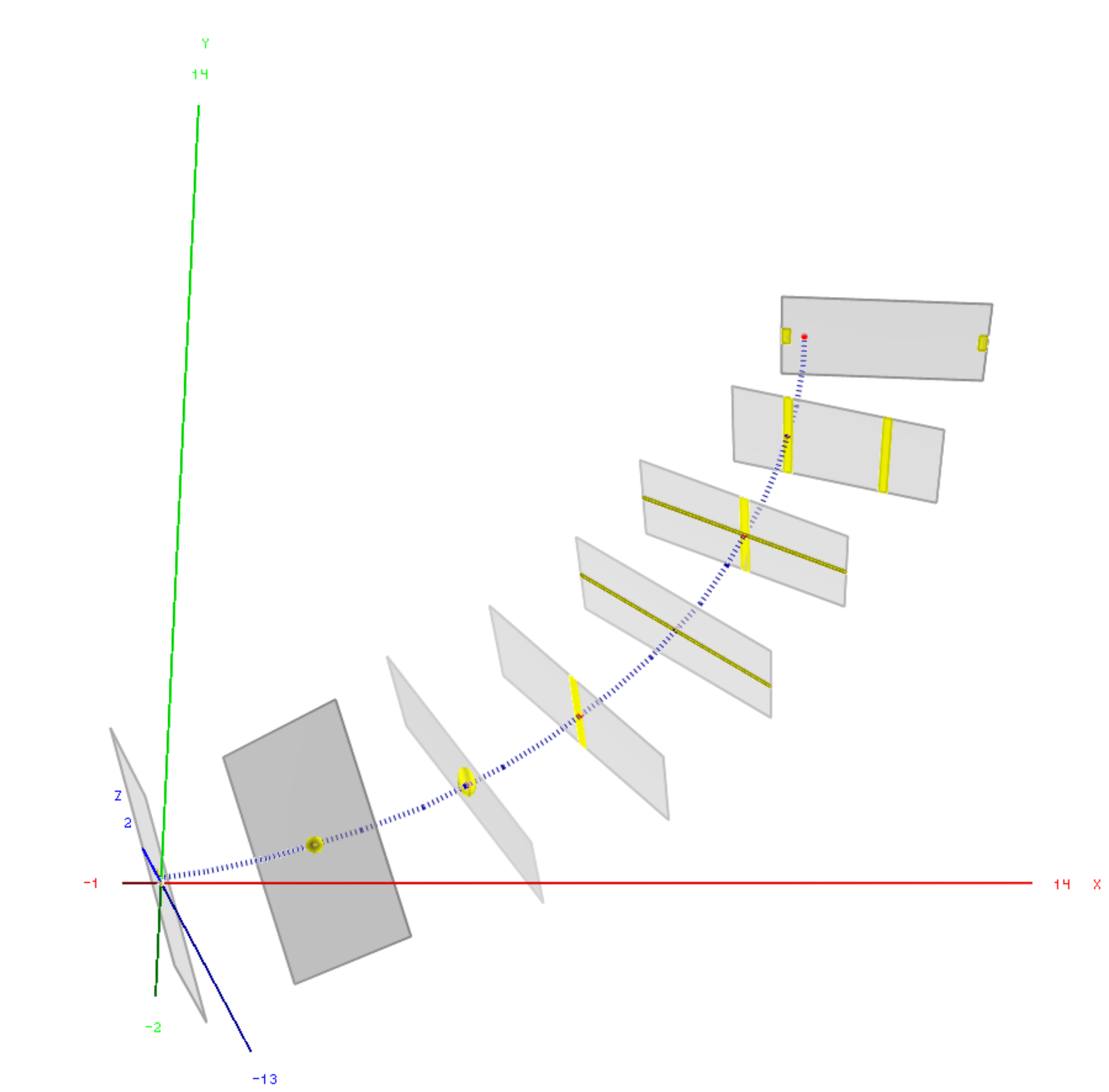}}\quad
        \subfigure[Forward (cyan) and backward (magenta) fit with covariances of the state updates.]{\includegraphics[trim =  16cm 2.25cm 20cm 4.6cm, clip, width=0.4\textwidth]{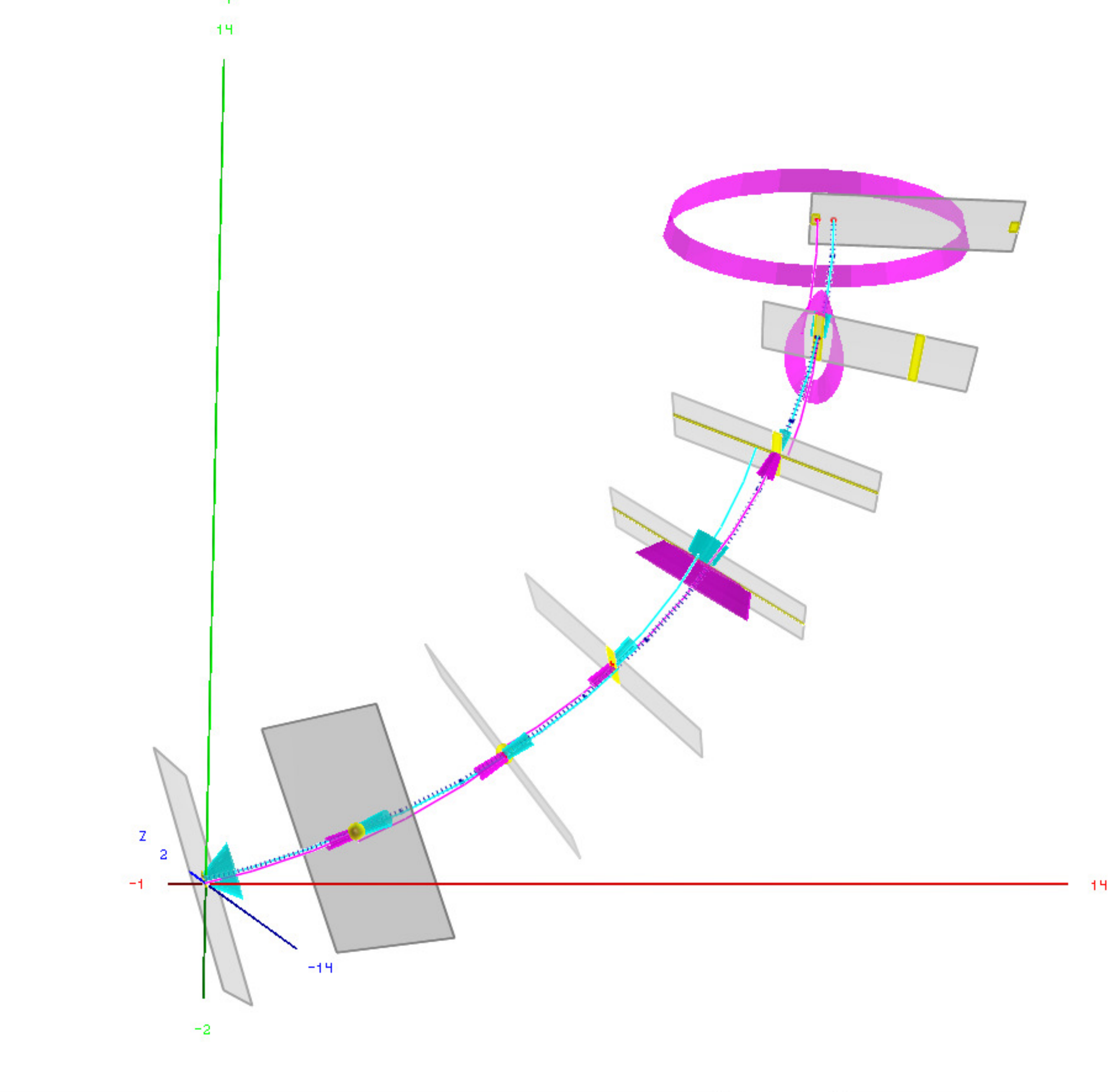}}\quad
        \subfigure[Smoothed track with covariance (blue).]{\includegraphics[trim =  16cm 2.25cm 20cm 4.6cm, clip, width=0.4\textwidth]{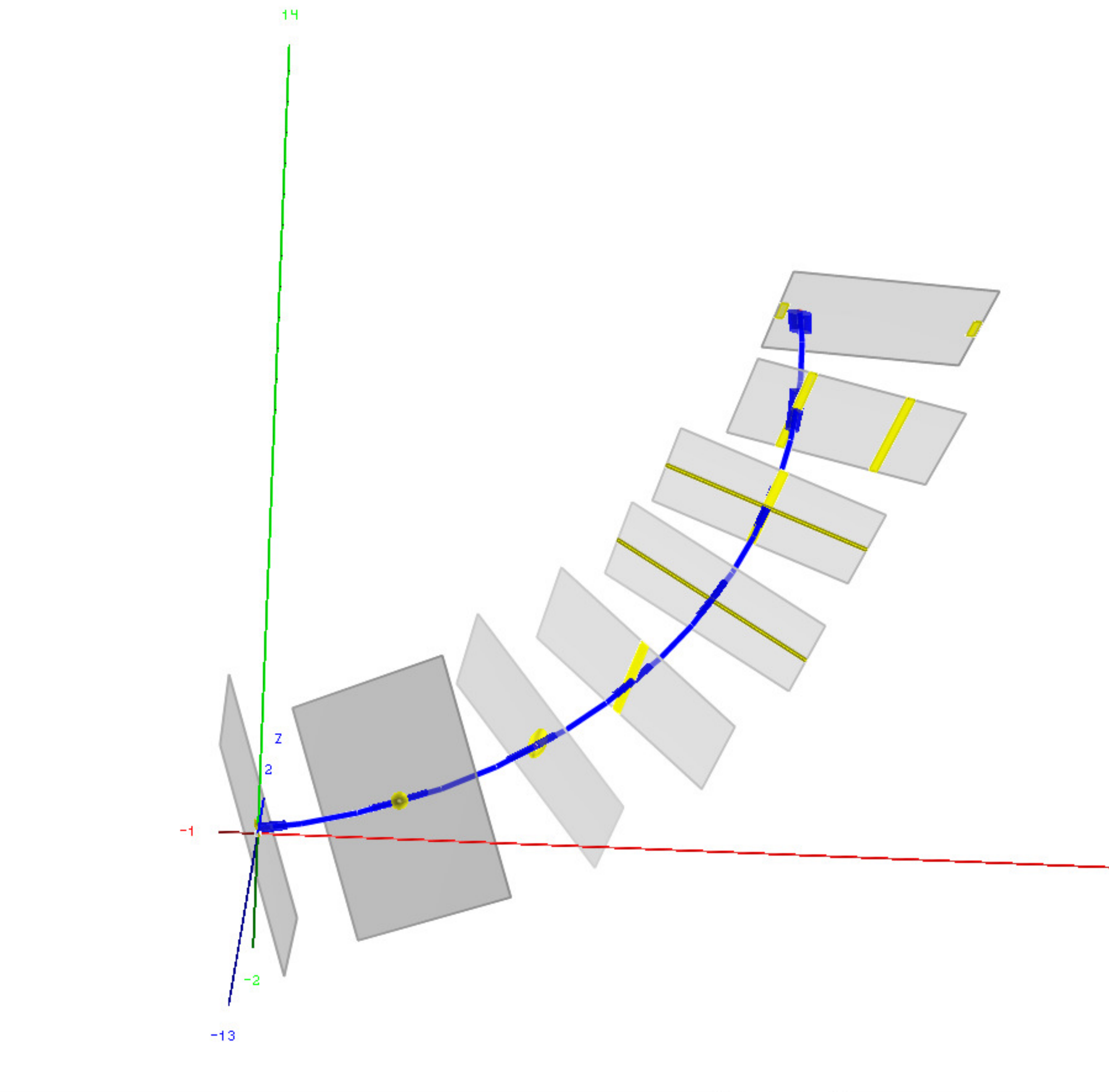}}
        \caption{\label{fig:eventDisplay}\genfit~event display screenshots~\cite{jojo}. The fit of a set of hits with the Kalman filter with reference track is shown. For demonstration purposes, the different hit types supported by \genfit~are used, starting from the origin: planar pixel hit, spacepoint hit, prolate spacepoint hit, two perpendicular planar strip hits, double sided planar strip hit, wire hit, wire hit with second coordinate measurement.}
\end{figure}

\genfit~features a sophisticated 3D event display, which allows to visualize fitted tracks. Detector geometry, measurements, detector planes, reference tracks, forward and backward fits (predictions and updates), smoothed tracks, and covariances of measurements and tracks can be drawn. Tracks can be refitted with different algorithms and settings, and fit results can be viewed instantly. Examples have been shown in Figs.~\ref{picgf2}-\ref{fig:eventDisplay}.
\clearpage
\clearpage
\begin{figure*}[ht] 
  \begin{center}
    \includegraphics[width=32pc]{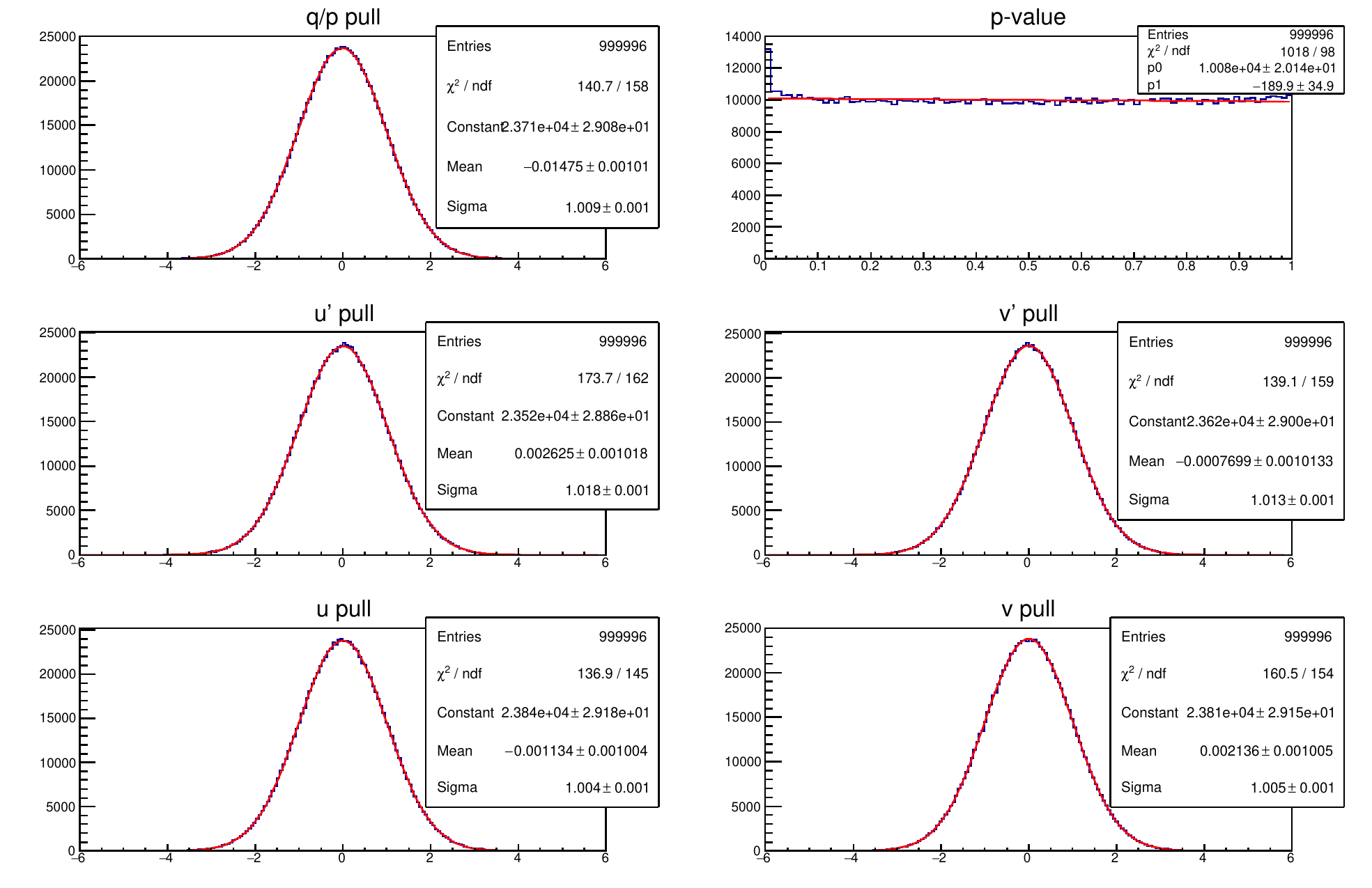}
    \caption{\label{standalonegf2}  Pull of the track parameters ($u,v,u$'$,v$' and $q/p$), with the pion momentum~= 1 GeV/$c$. Gaussian fits to these pull distributions are consistent with the expectation of a normal distribution.}
  \end{center}
\end{figure*}
\begin{figure*}[h!] 
  \begin{center}
    \includegraphics[width=32pc]{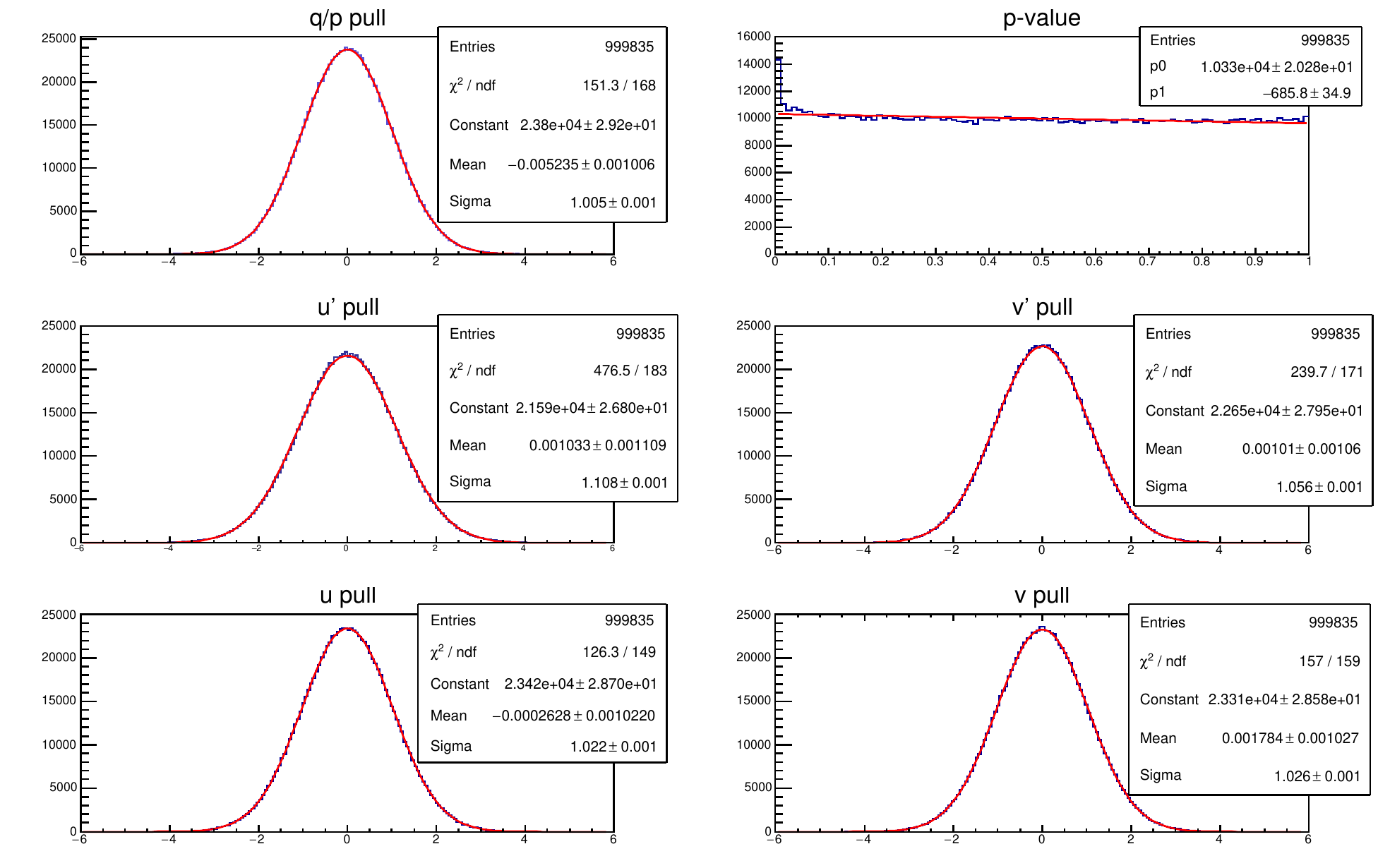}
    \caption{\label{standalonegf22}  Pull of the track parameters ($u,v,u$'$,v$' and $q/p$), with the pion momentum~= 100 MeV/$c$. Gaussian fits to these pull distributions are consistent with the expectation of a normal distribution.}
  \end{center}
\end{figure*}
\begin{figure*}[h!] 
\begin{center}
  \includegraphics[width=32pc]{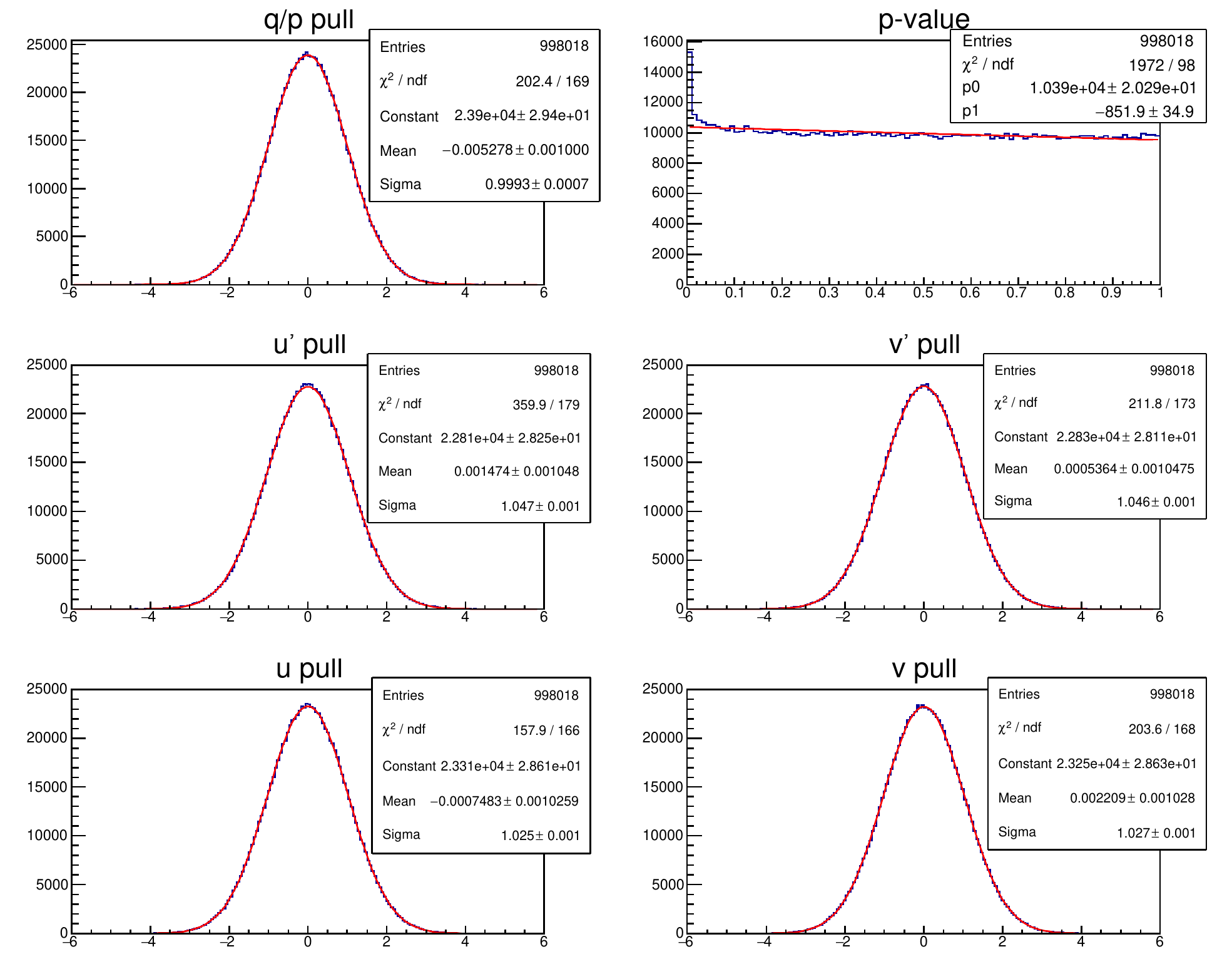}
    \caption{\label{standalonegf222}  Pull of the track parameters ($u,v,u$'$,v$' and $q/p$), with the pion momentum~= 50 MeV/$c$. Gaussian fits to these pull distributions are consistent with the expectation of a normal distribution.}
\end{center}
\end{figure*}
\clearpage
\subsection{Performance of \genfit~in basf2}
\label{sec-5}
\begin{figure*}[ht] 
  \begin{center}
    \mbox{
      \includegraphics[width=24pc]{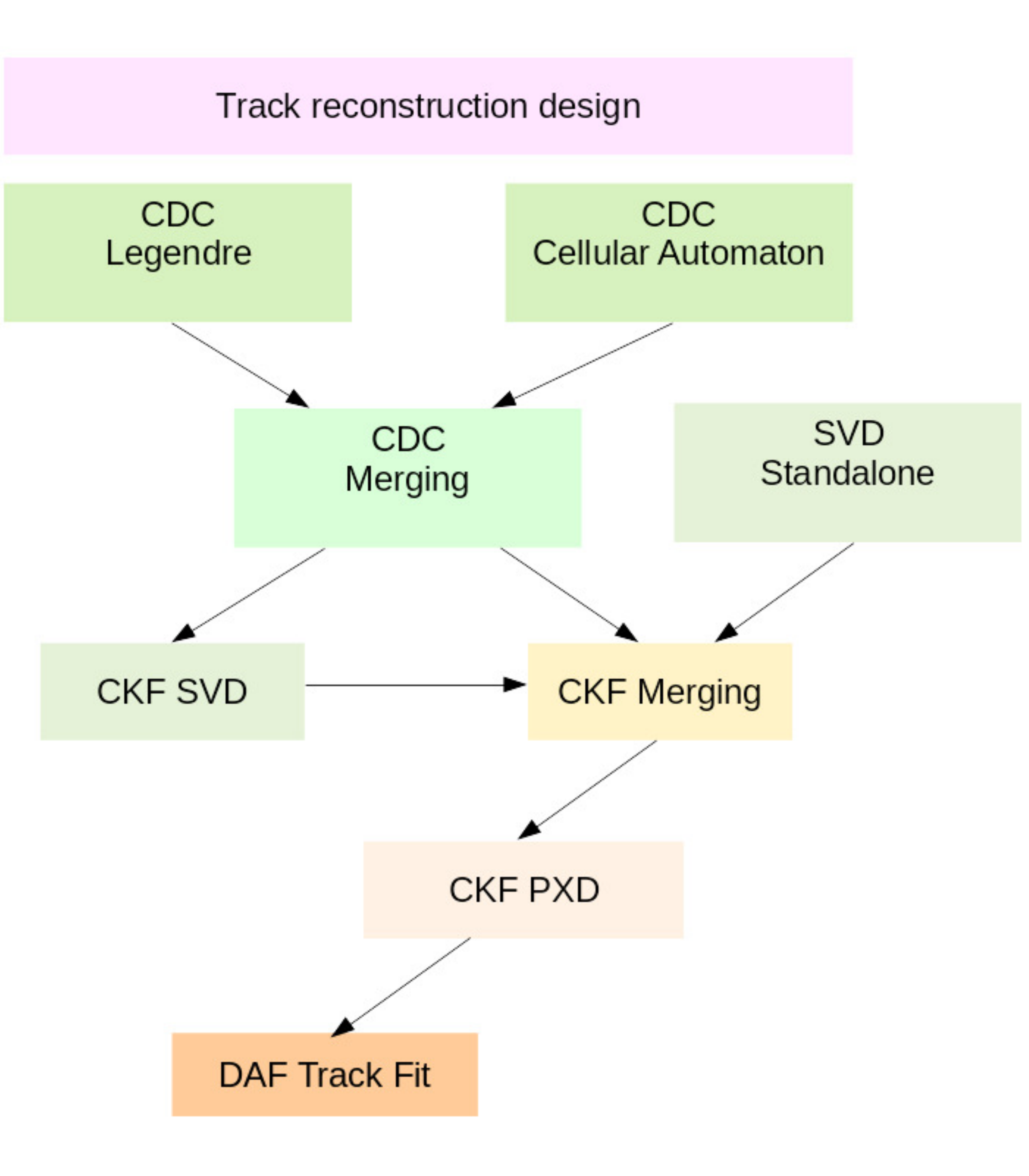}\quad
    }
    \caption{\label{fig10} Scheme of the track reconstruction design in basf2~\cite{stespa}.}
  \end{center}
\end{figure*}
\genfit~performance depends on many parameters, such as computer hardware and compiler, number and types of measurements, momentum, geometry, magnetic field, number of iterations, convergence settings, and so on.
Nevertheless, a study has been conducted in both, basf2 and PandaRoot, to give the reader an idea of the performance expected in high energy physics using \genfit. Figure~\ref{fig10} shows the track reconstruction design in Belle II. The Belle II detector is structured with an internal  vertex detector (VXD), which is made by 2 inner layers of silicon Pixel Detector (PXD) and 4 double-side  strips of Silicon Detector (SVD); then a Central Drift Chamber (CDC) is given for tracking purposes and particle identification.  A time-of-propagation (TOP) detector is also furnished, surrounded by an electro-magnetic calorimeter (EM), while the outer detector is for $\mu$ and $K_L^0$ detections. A Combinatorial Kalman Filter (CKF) was implemented at software level for track finding issues. \genfit~takes care of track fitting in basf2: it acts in a way to bind together track segments and information of PXD (planar hits), SVD (position along strips), and CDC (wire  and drift time).
The execution time of the basf2 {\tt GenFitter} module has been measured on a current 3.4 GHz office PC in single threaded operation, an Intel machine with 4 cores. 
All code has been compiled with {\tt -O3} optimization settings.

The module performs the fitting and a few more other features, like producing the \genfit~track from a {\tt TrackCand}, storing the fitted track in an output array, creating relations etc., resulting in an overhead of a little less than 1 ms. Tracks are generated with a particle gun, with $\theta = 100^o$ and a momentum of 0.9 GeV/$c$ in a constant magnetic field equal to 1.5~T. The resulting tracks have $72 \pm 2$ {\tt TrackPoints}. \genfit~is configured to do from 3 to 10 iterations with default convergence criteria. From the results shown in Tab.~\ref{tab:performance}, one can see that the {\it Kalman Fitter with reference track} needs less iterations to converge. Material lookup takes around 2.2 ms per iteration for the Kalman filter. The {\it Reference Kalman fitter} is also faster here, since the reference states are not recalculated if they are close to the smoothed states of the previous iteration.
This is also the reason why the {\it Reference Kalman fitter} can often finish after 2 iterations. The reference track is already so close to the smoothed track, that it would not change in subsequent iterations. Table~\ref{tab:performance} shows that the calculation of the material effects takes a large proportion of the computing time (of course, it strongly depends
on the complexity of the detector geometry under exam), and the Kalman fitter looks reasonably fast.

\begin{table*}[!htb]
\caption{ Execution time of the GenFitter module in basf2.}
\begin{center}
\begin{tabular}{lrccccl}
\hline\hline
Fitter & no material effects & with material effects & n.~iterations\cr \hline \hline
\noalign{\vskip2pt}
   Kalman    & 3.4 ms & 10 ms & 3 \cr
    Reference Kalman & 4.0 ms & 8.2 ms & 2 \cr
    DAF              & 7.0 ms & 11 ms & 4 \cr \hline
\end{tabular}
\end{center}
\label{tab:performance}
\end{table*}

\begin{figure*}[ht] 
  \begin{center}
    \mbox{
     \hspace{-1.2 cm} \subfigure[]{\scalebox{0.27}{\includegraphics{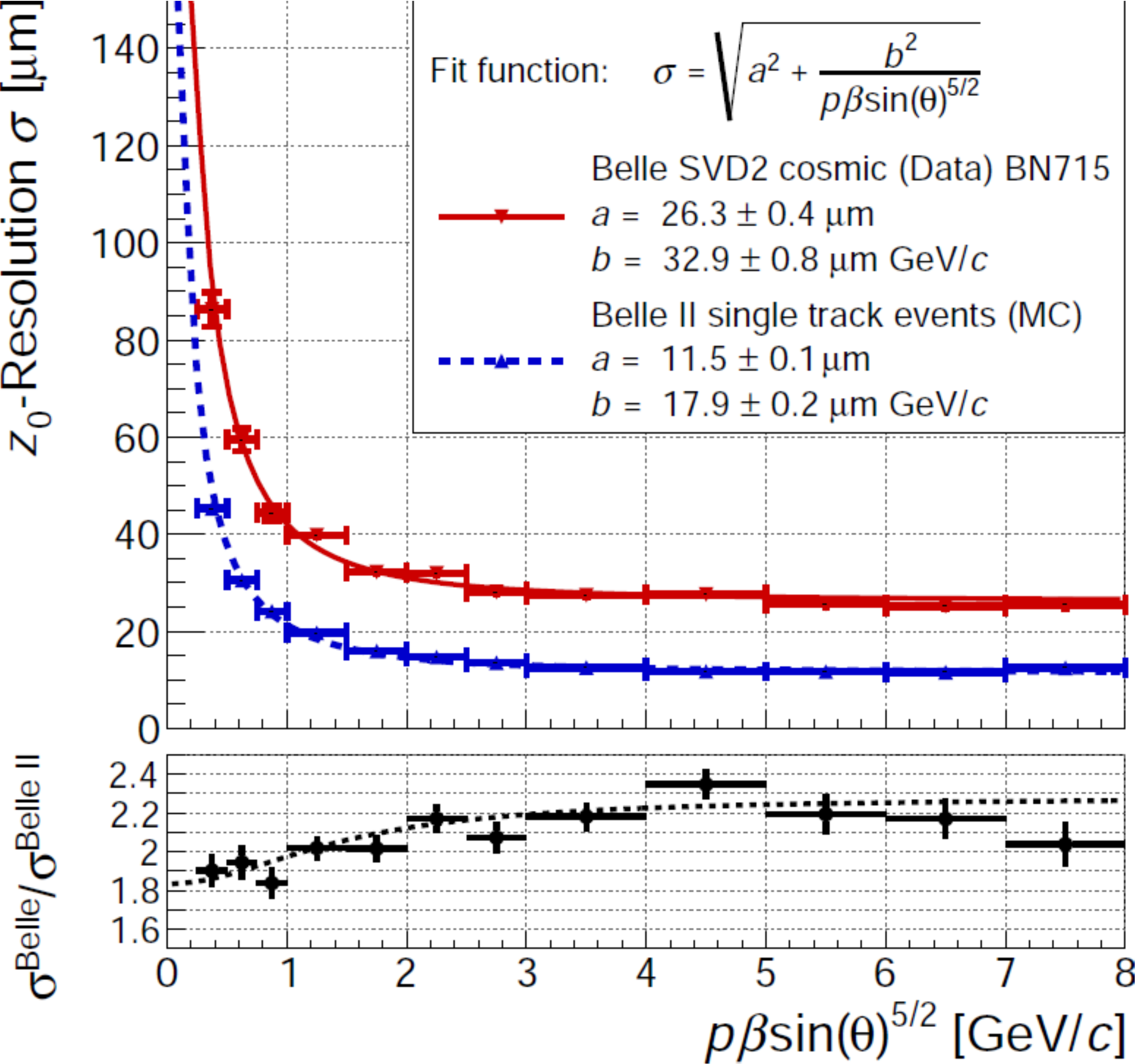}}}\quad \hspace{1 cm}
     \hspace{1.0 cm}  \subfigure[]{\scalebox{0.27}{\includegraphics{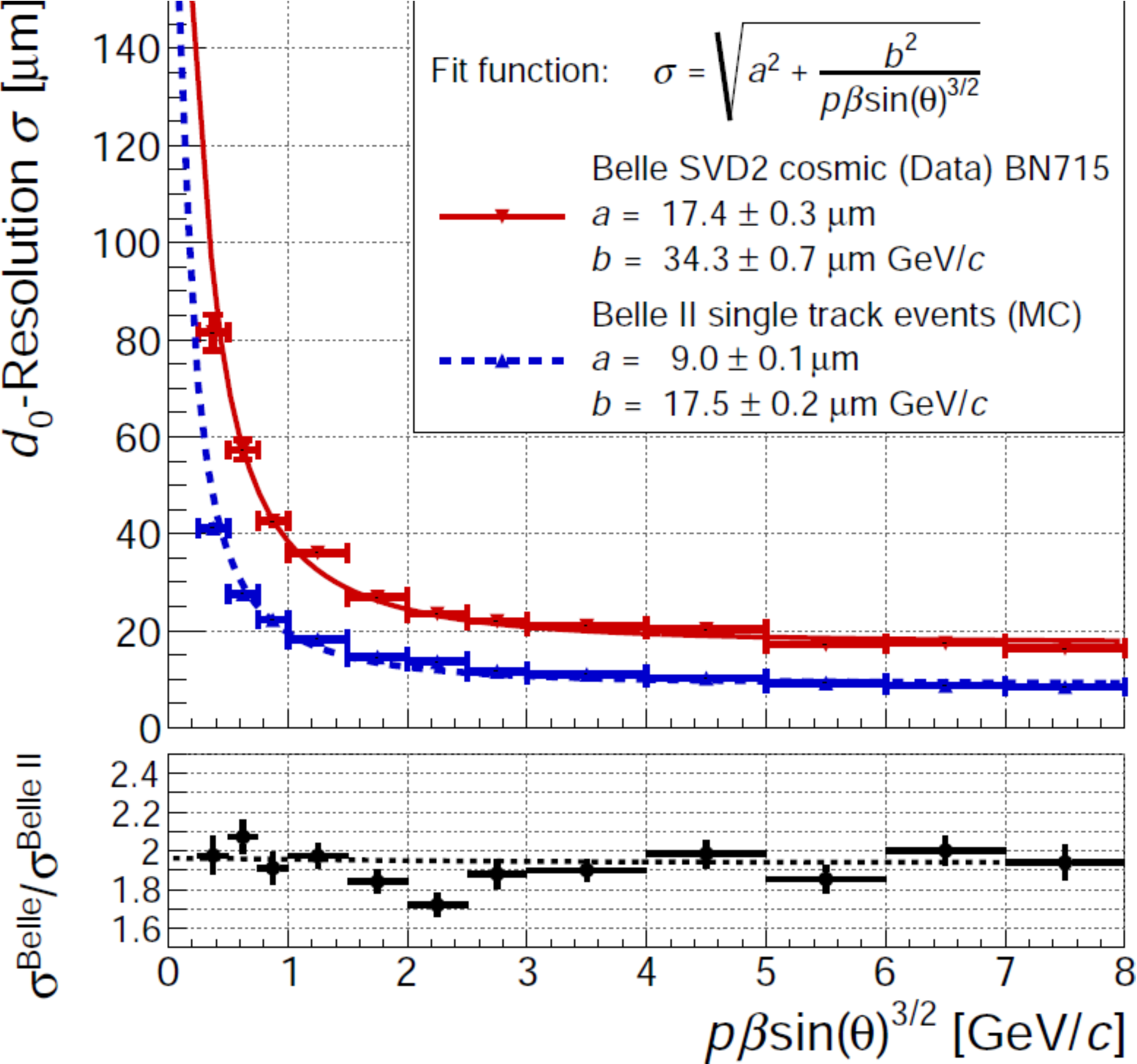}}}
    }
    \mbox{
     
      \hspace{-4.5 cm}\subfigure[]{\scalebox{0.37}{\includegraphics{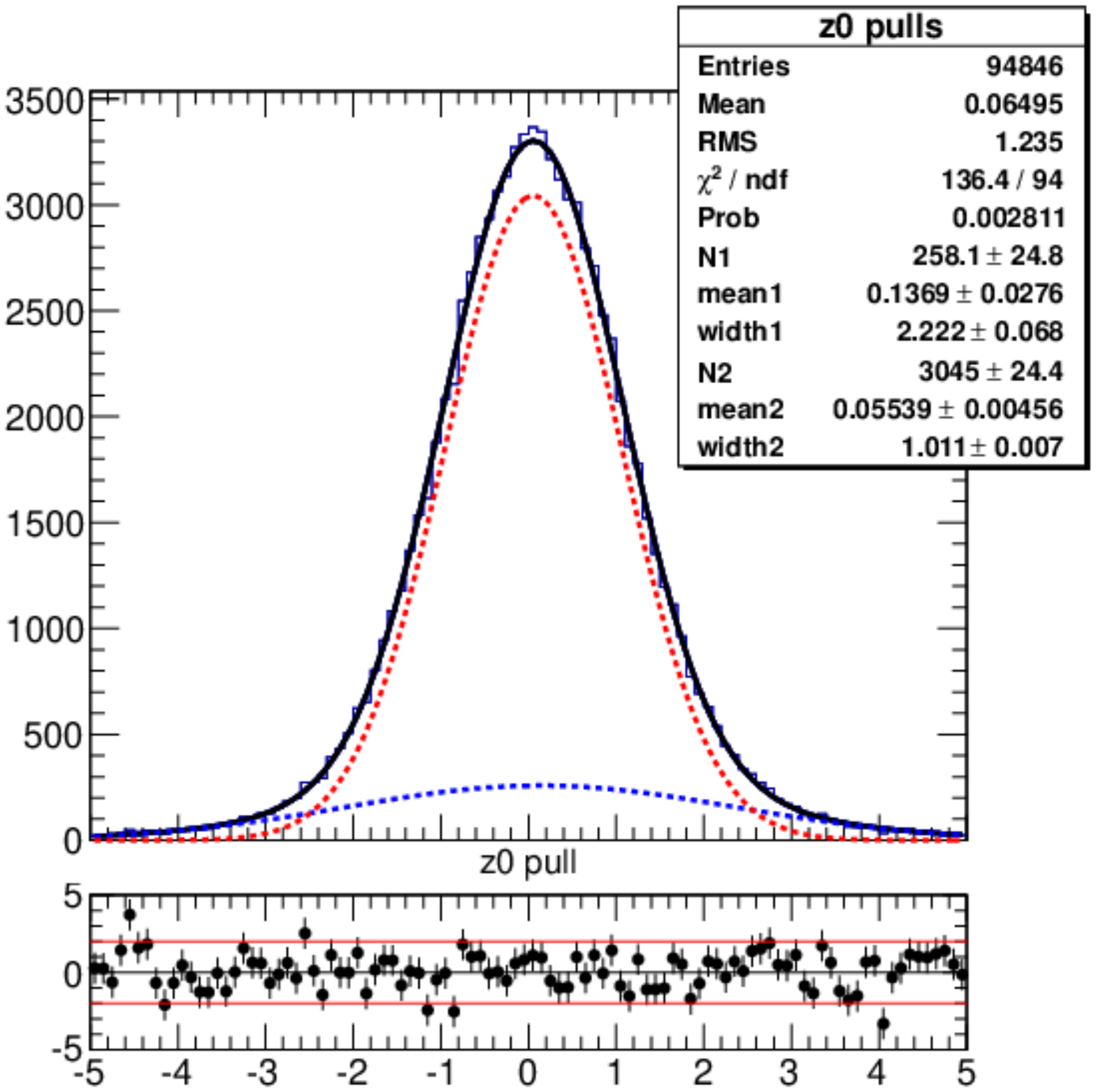}}}\quad
      \subfigure[]{\scalebox{0.37}{\includegraphics{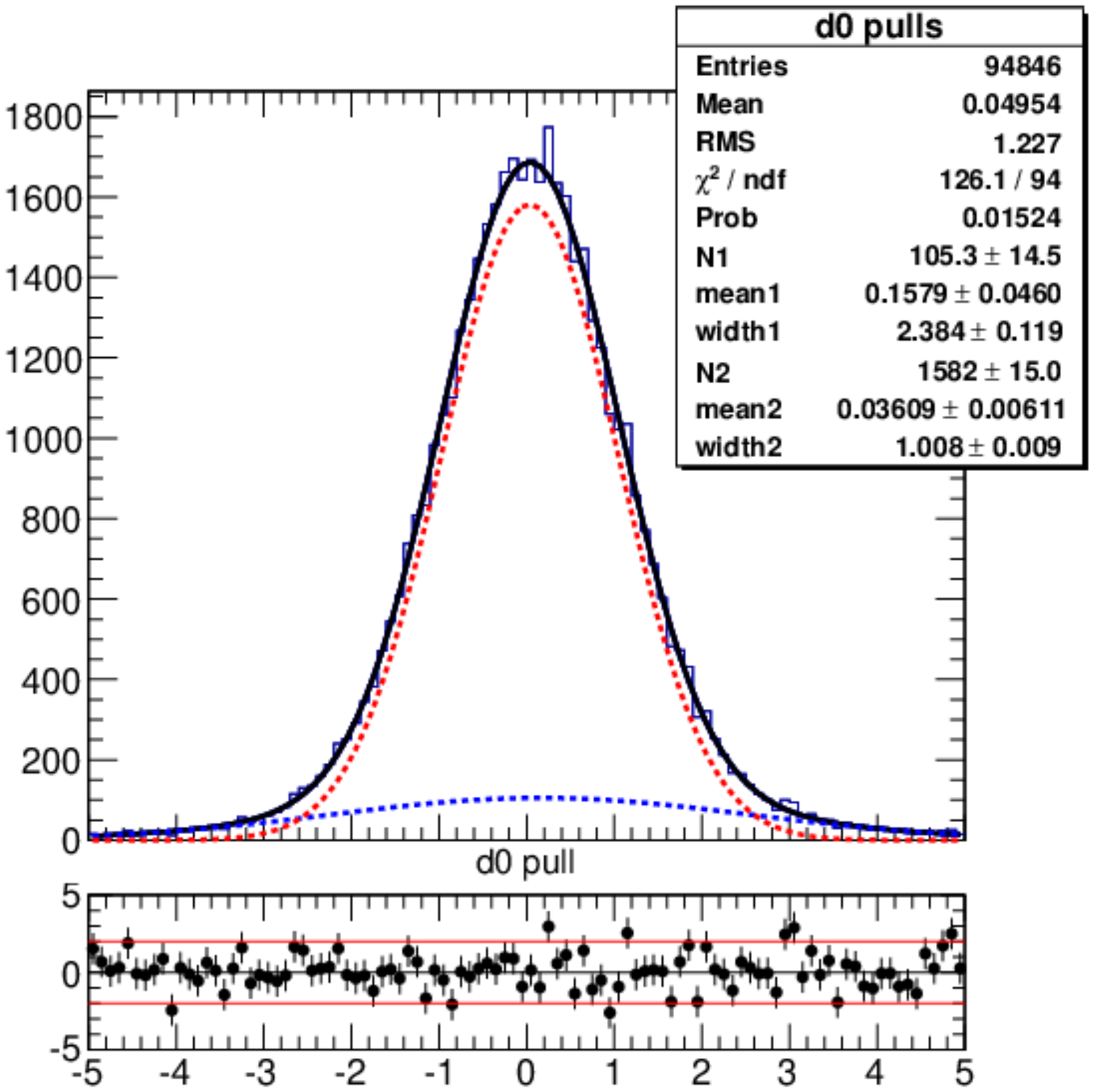}}}
    }
    \caption{\label{d0z0belle} Study of track reconstruction in Belle II, using cosmics collected during February-March 2018: $z_0$ (a) and $d_0$ (b) resolution vs momentum. $z_0$ and $d_0$ are defined as the projections of the z axis of the POCA, and that of the x-y plane of the POCA, respectively. Fits to real data (cosmics, red plot) and to the MC simulated single track events (blue plot) are displayed. Figures (c) and (d) show respectively the fits to the pull distributions of those parameters~\cite{PB1}-\cite{elicharm2}.}
\end{center}
\end{figure*}

\begin{center}
\begin{figure*}[ht] 
      \includegraphics[width=28pc]{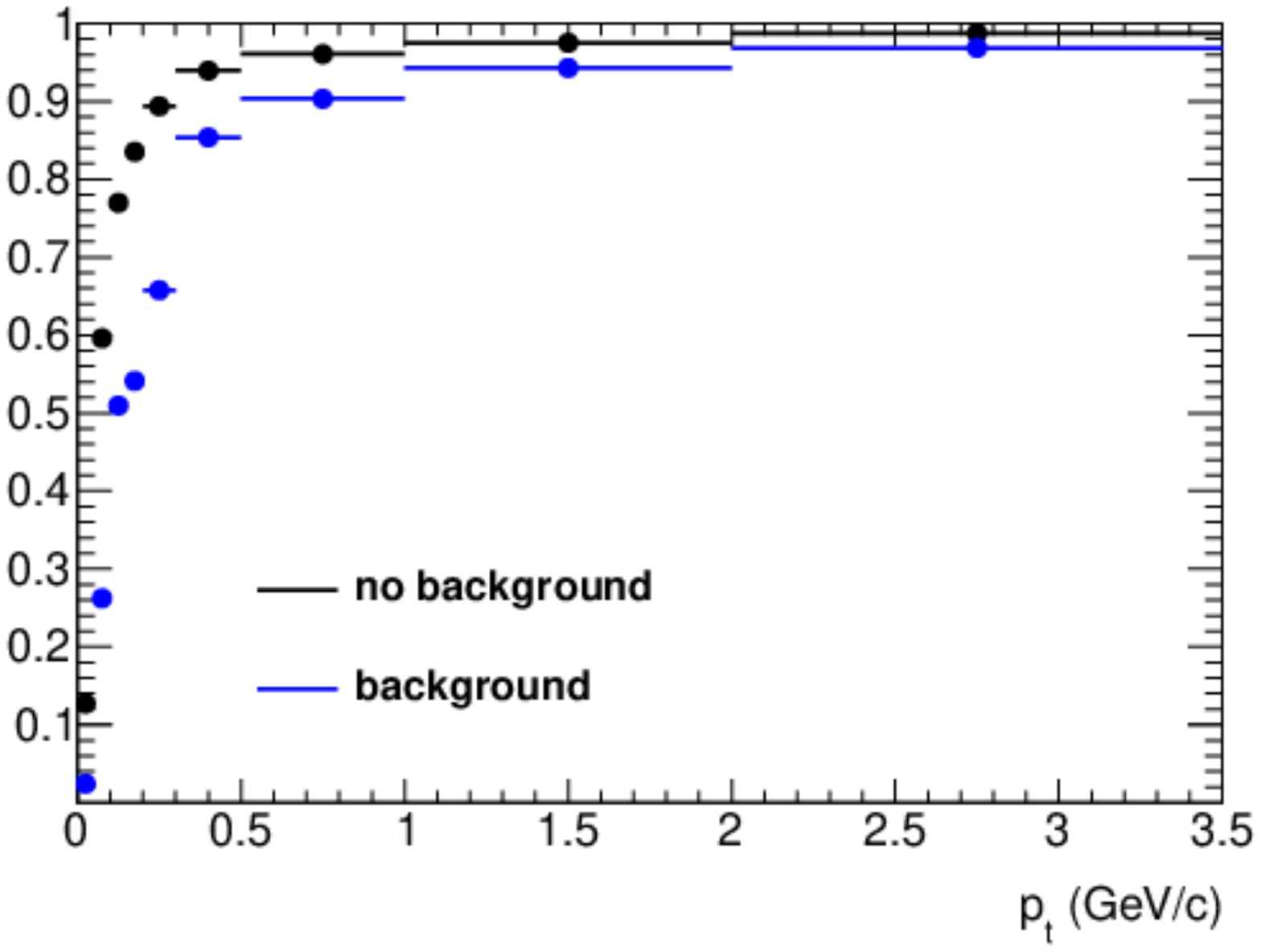}
    \caption{\label{effibelle2} Efficiency of charged particles reconstructed with \genfit~in basf2~\cite{PB1}, with and without machine background simulation.}
  \end{figure*}
\end{center}

Examples of the good performance of \genfit~in basf2 are shown in Figs.~\ref{d0z0belle}-\ref{effibelle2}: in the former a study with cosmics is performed, showing the good shape of the track reconstruction software code; in the latter a study showing the efficiency of tracks, reconstructed with basf2, is shown. The pull distributions of those parameters, defined as the ratio between the difference of reconstructed and generated values, and the error of such parameter, is parameterized by a double Gaussian function, and displayed in Fig.~\ref{d0z0belle}.  
\clearpage
\subsection{Performance of \genfit~in PandaRoot}

\begin{figure*}[ht] 
\begin{center}
\includegraphics[width=25pc]{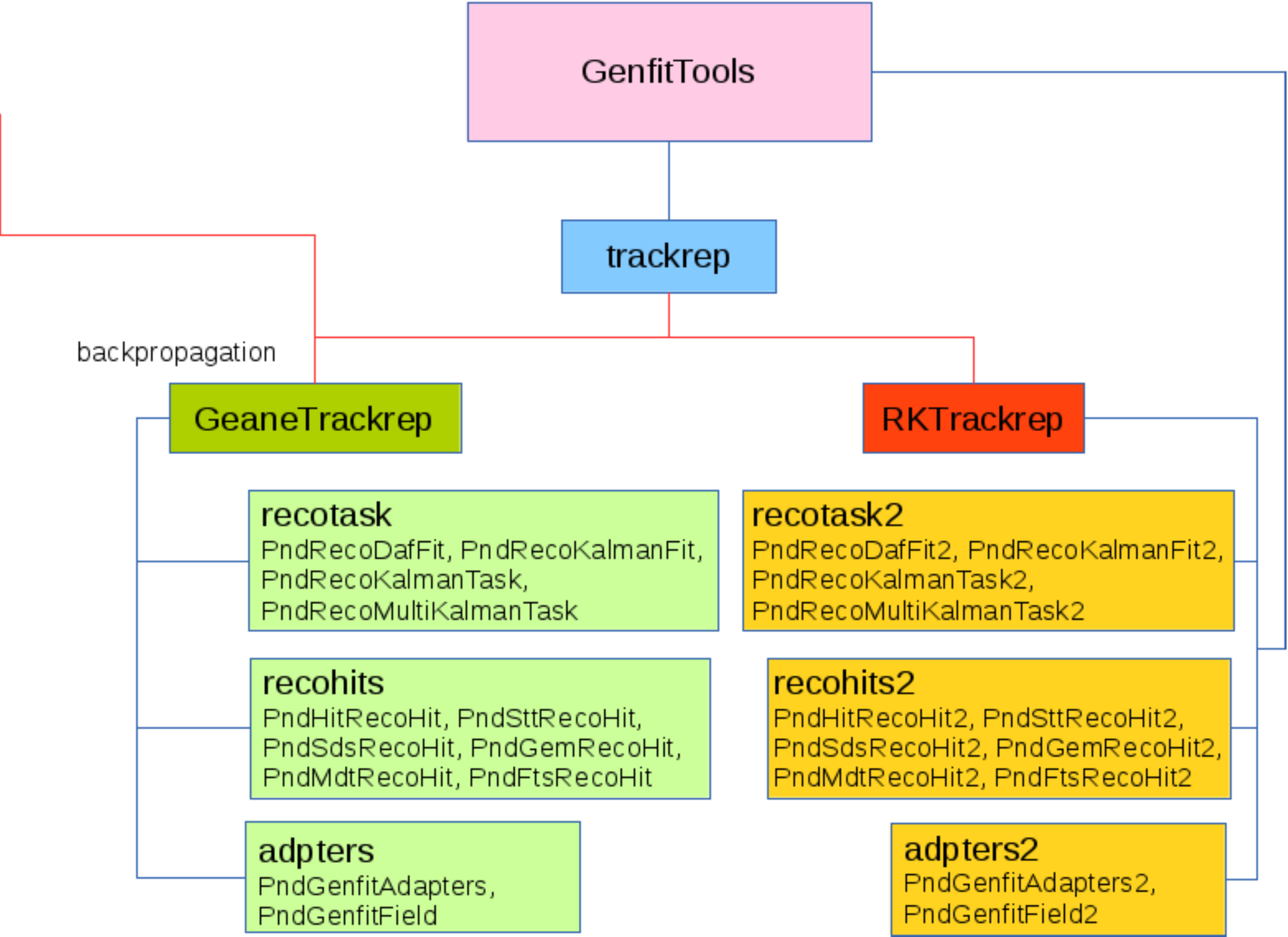}
\caption{\label{fig4}  Structure of the {\tt GenfitTools} package, in the PandaRoot trunk 28747~\cite{elichep}.}
\end{center}
\end{figure*}

An interface has been built between \genfit~and PandaRoot: {\tt GenfitTools}. The basic structure of {\tt GenfitTools} is the same used to interface the previous \genfitone~package, but with a few modifications due to the new classes introduced in \genfit~(see Fig.~\ref{fig4}).
The new {\tt GenfitTools} structure has been created to allow both, \genfitone~and \genfit, to be used as fitting tool, and then compare their results.

{\tt GenfitTools}  includes several objects: \emph{adapters}(2); \emph{recohits}(2); \emph{recotasks}(2); \emph{trackrep}.

The objects \emph{adapters}, \emph{recohits}, \emph{recotasks} refer to the \genfitone~interface to the PandaRoot classes, and the  objects \emph{adapters}2,  \emph{recohits}2, \emph{recotasks}2 refer to the \genfit~interface to the PandaRoot classes.  \emph{PndTrack} is the standard object for PandaRoot tracks, and it is converted into a \genfit~or \genfitone~{\tt Track} by means of the \emph{adapters}(2), which are able to deal with changes in the \genfit~code without changing additional packages.

\begin{figure}[ht]
  \centering
  \mbox{
    \subfigure[]{\includegraphics[width=27pc]{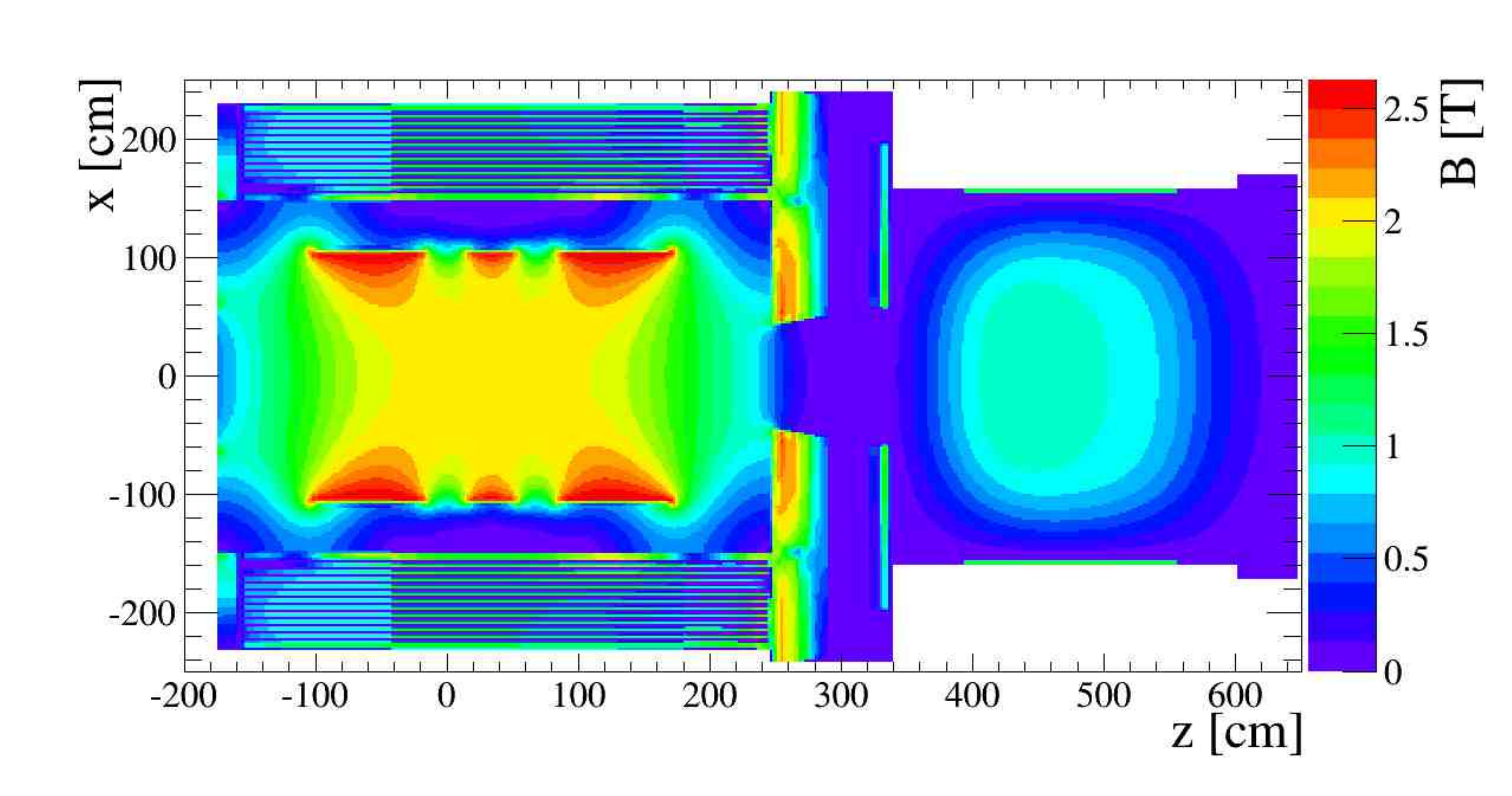}}}
  \mbox{
 \hspace{1 cm}\subfigure[]{\includegraphics[width=25pc]{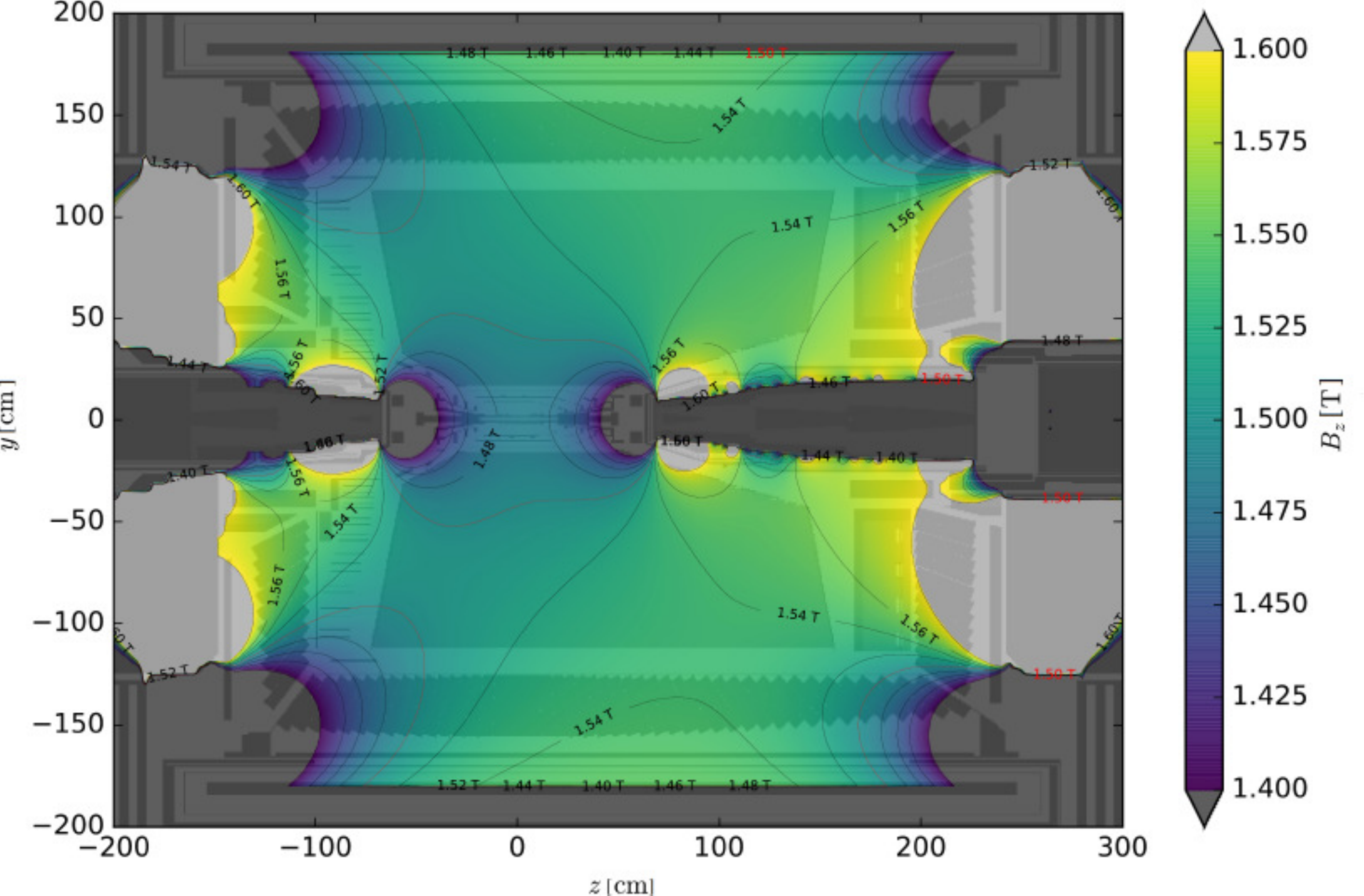}}
}   
\caption{\label{bfield} (a) Magnetic field scheme in \PANDA~\cite{elichep}: in the central part of the detector it is constant and equal to 2T; in the dipole area the maximum bending power is 2 T$\cdot$m. (b) Magnetic field scheme in Belle II~\cite{PB3}: it is a solenoidal field of 1.5~T at the center of the detector.}
\end{figure}

    In order to test the performance of the \genfit~and the new {\tt GenfitTools} package,  a so-called particle-gun generator has been used. Five particle hypoteses have been tested: electrons, muons, pions, kaons and protons. A MC particle-gun  generator is used to produce single particle beam for each different particle hypotheses, with different $p_T$ values, running at the same antiproton beam momentum\footnote{A change in the antiproton momentum beam changes the flux density  of the dipole magnetic field in the \PANDA~forward spectrometer.}. For the example mentioned in this section a beam momentum of 15 GeV/c was used. We tested particles emitted at a fixed polar angle ($\theta$ = 60$^0$). We also checked the \genfit~performance in PandaRoot by changing the beam momentum, to evaluate the resolution and pull of the tracking parameters. The magnetic field map used in these tests is a realistic one, as shown in Fig.~\ref{bfield}(a). Since the \PANDA~detector is composed by two spectrometers, one with a solenoidal field (on the left part of Fig.~\ref{bfield}(a))  and another with dipole field (on the right of Fig.~\ref{bfield}(a)), the intermediate region is characterised by strong inhomogeneities which has to be dealt by the extrapolation and fitting code. For comparison,  the Belle II magnetic field scheme is shown in Fig.~\ref{bfield}(b). More details related to this study are published in Ref.~\cite{elichep}.

The track representation used in the following tests is not {\tt RKTrackRep}, because PandaRoot currently uses GEANE~\cite{geane1}; therefore an estimate on how long it takes to reconstruct an event with \genfit~in PandaRoot framework is biased, and not comparable with the numbers measured for basf2.

%
\begin{figure*}[ht]
\label{eliplot}  
\begin{center}
\includegraphics[width=23pc]{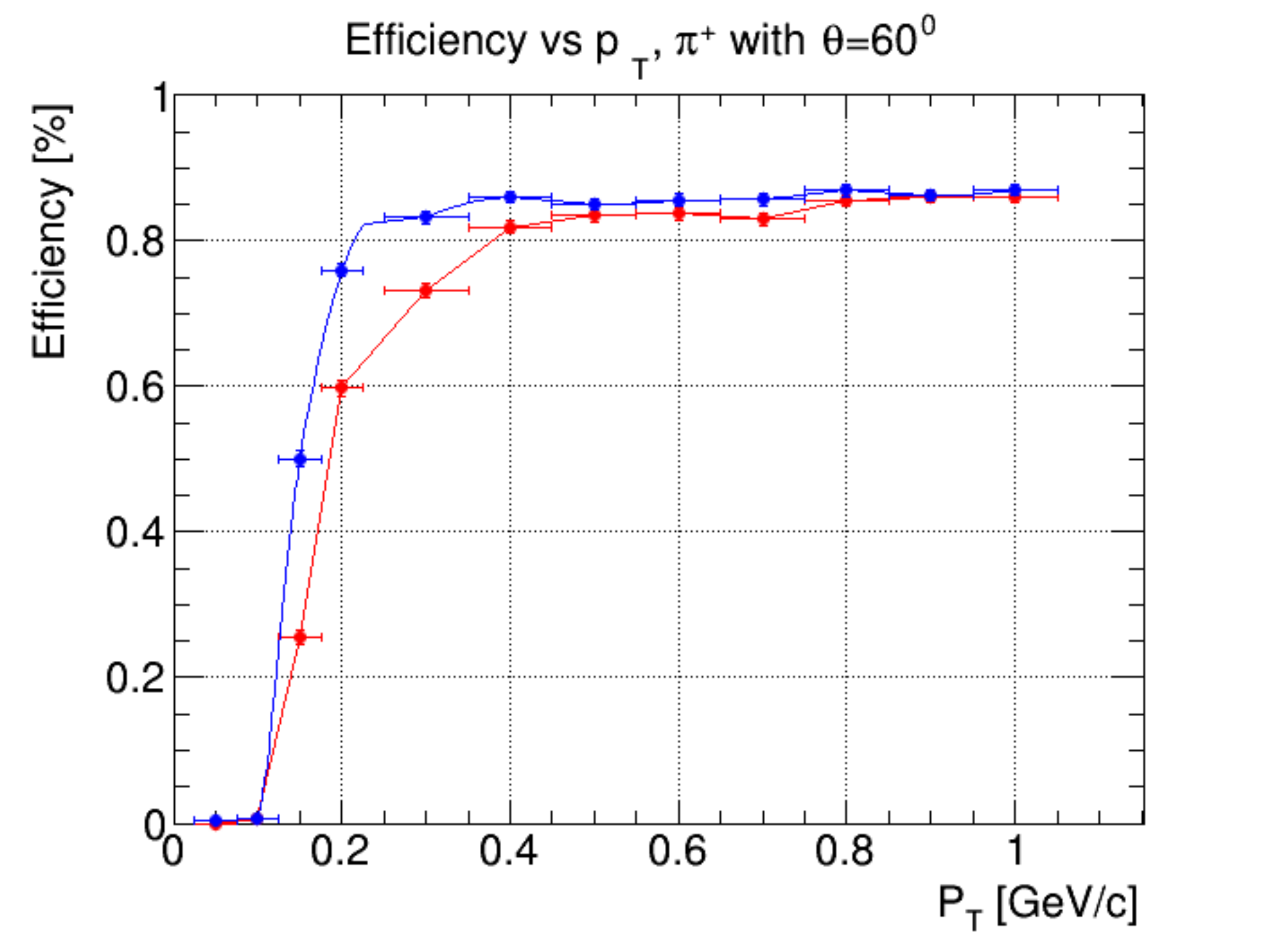}
\caption{\label{plotPB} Efficiency versus the transverse momentum ($p_T$) of charged pions, emitted at the polar angle value $\theta$ = 60$^0$, in PandaRoot. A comparison between the old tool (red squares) and the new \genfit~tool (blue points) is provided.  The track fitting efficiency is evaluated in a 3$\sigma$ window around the reconstructed momentum~\cite{elichep}.}
\end{center}
\end{figure*}
\begin{figure}[ht]
  \centering
  \mbox{
    \subfigure[]{\scalebox{0.37}{\includegraphics{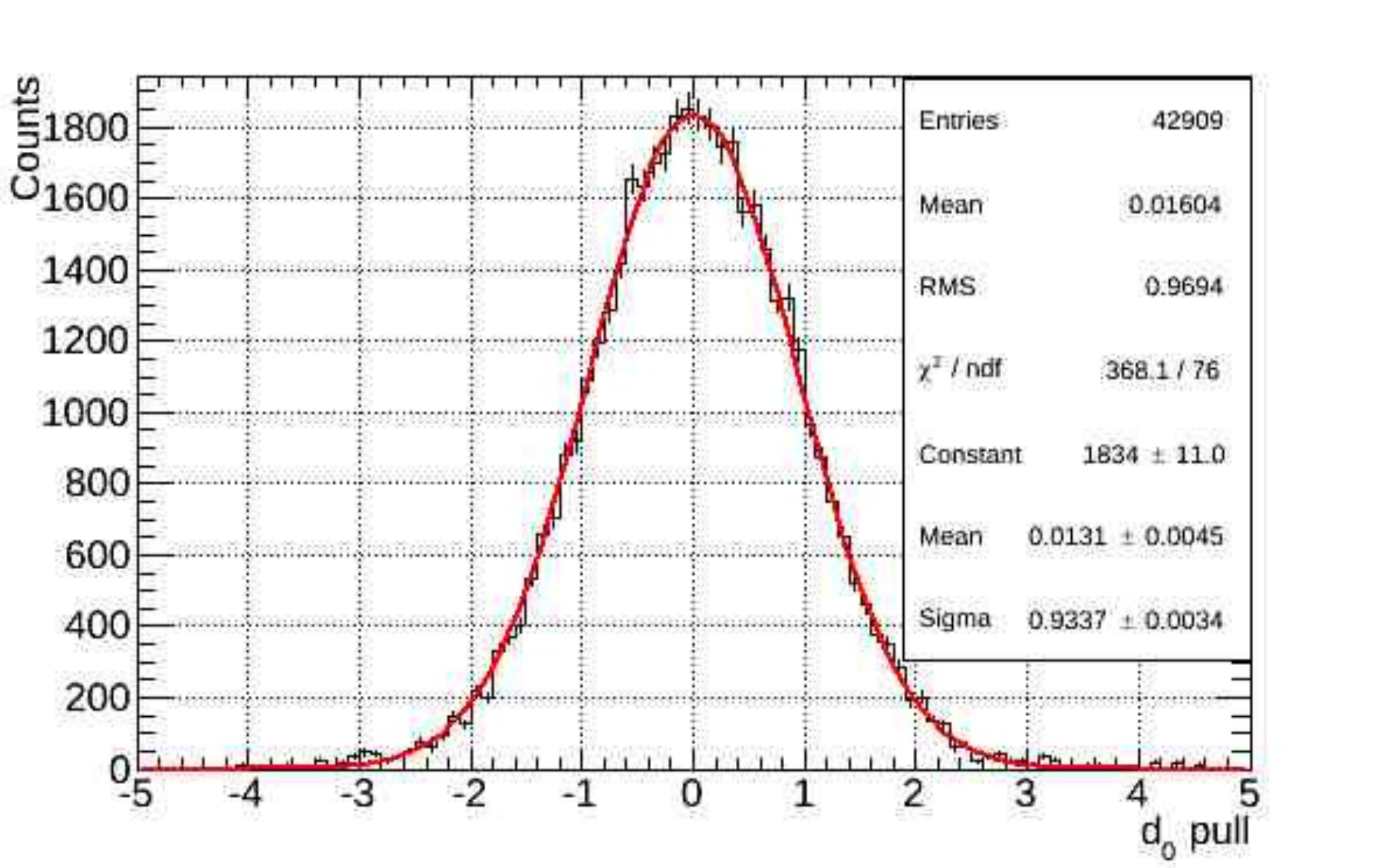}}}\quad
    \subfigure[]{\scalebox{0.37}{\includegraphics{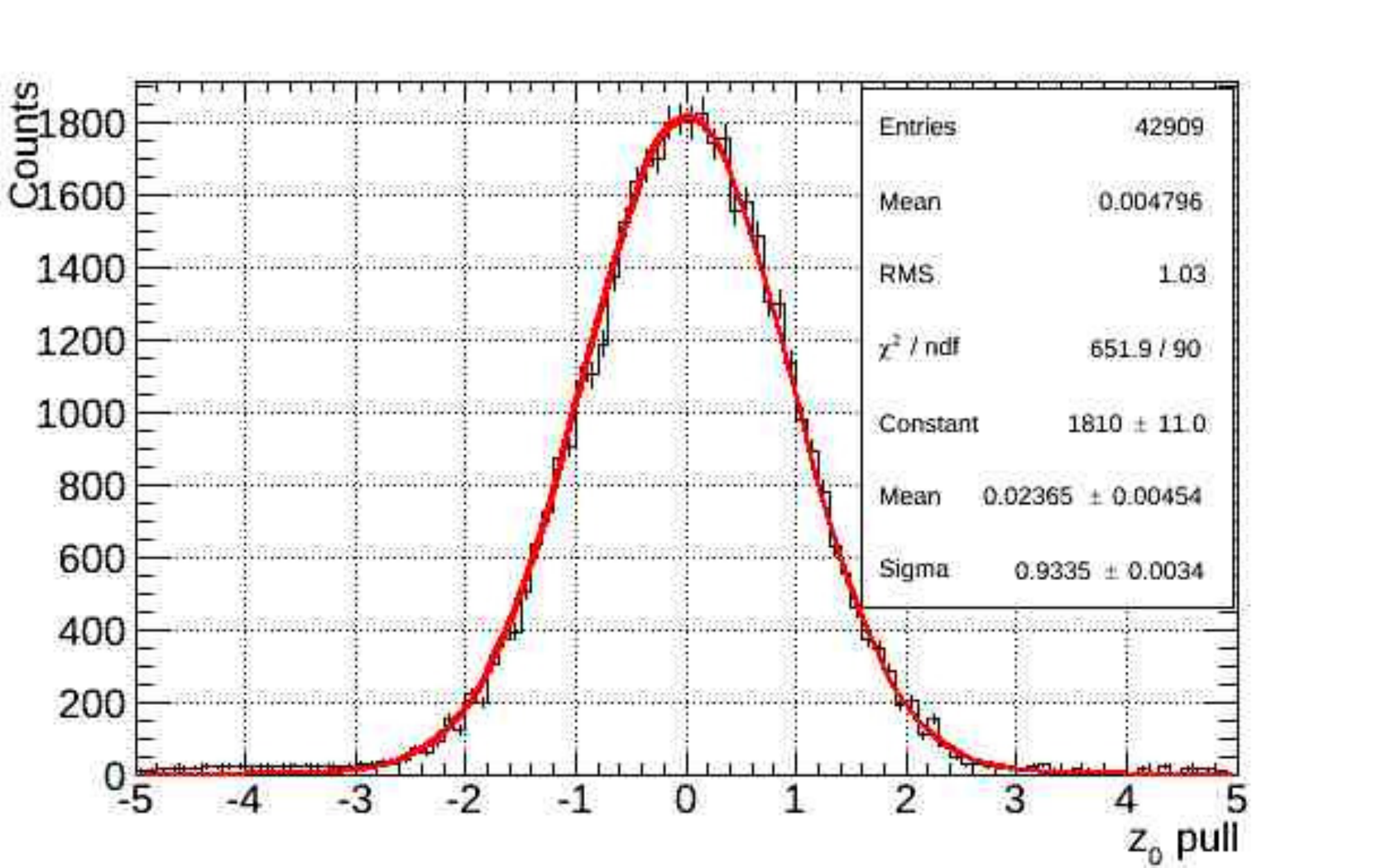}}}
}
\caption{ \label{fig60} Generated values are subtracted from the reconstructed values of the variables \textit{$d_0$} and \textit{$z_0$}, and then divided by their errors. These plots have been obtained for \textit{$p_T$} = 400 MeV/\textit{c} and they  show the pull distributions~\cite{elichep}  using \genfit~in PandaRoot. The width of the Gaussian fit to both  distributions is consistent with being 1.0, as expected for a correct pull distribution, and the mean is consistent with being zero.}
\end{figure}

Figure~\ref{plotPB} shows the performance of tracking tools in 2 different  PandaRoot trunk revisions: the old tool (red squares) in comparison with the new \genfit~tool (blue squares). The antiproton beam momentum is set to 15 GeV/c.   No significant differences  are observed  in the 2 different PandaRoot trunk revisions under investigation, for high $p_T$; however, for $p_T <$0.35 GeV/$c$ the \genfit~performance is now better. In the PandaRoot trunk revision 28747 a switch is provided to use both, \genfit~and \genfitone. \genfit~has been introduced in the official PandaRoot releases since 2017.

The result of our simulation shows  a clear improvement for reconstruction of low momentum tracks (see Fig.~\ref{plotPB}). A minimum transverse momentum threshold of 50 MeV/c is imposed, since that is required for at least one hit in the Straw-Tube-Tracker (STT).
\genfit~is built in a way to track low momentum tracks down 0 MeV/c.  

We summarize in Fig.~\ref{fig60} a preliminary achievement of this work in PandaRoot: it shows the pull distribution of the spatial tracking parameters $d_0$ and $z_0$, running the particle-gun beam at $p_T$ = 1 GeV/$c$. As expected, the width of a Gaussian fit to the distribution is consistent with 1.0, and the mean with zero. Small variations are observed if we decrease the $p_T$ value, but they are not significant, and the results confirm the stability of the fitter~\cite{dots2016}.

\section{Conclusion}
In summary, we have demonstrated that \genfit~can successfully provide track-fitting tools when the magnetic field is nearly homeogeneous (such as in Belle II) and when it is non-homogeneous (such as in \PANDA). \genfit~has shown good track-fitting performance and has been validated with MC samples. It has successfully been integrated into PandaRoot and basf2. It easily adapts to different detector setups and geometries. With the new fitter implementation  classes, compared to the former \genfit, the performance  of low momentum track reconstrucion  is improved. Tests have been performed changing  particle hypotheses: electron, muon, pion, kaon, and proton.  The Runge-Kutta track representation works well in basf2, and is available also in PandaRoot now. The newly available interfaces to Millepede II and RAVE are also useful and easy to utilize, as it is the new event display feature included in the toolkit. Preliminary results are promising, as reported in Ref.~\cite{elicharm2} (cosmics, and first data collected during the early time of  Phase 2 of Belle II for a total integrated luminosity equal to 5 pb$^{-1}$). \genfit~is now used in several projects, including Belle II, COMET, \PANDA, FOPI, SHIP, FOOT and BGO-OD. We encourage scientists to make use of our generic tracking tool and participate in our ongoing work to further improve and extend \genfit.

\appendix

\section{Coordinate system}
\label{coordinates}

GENFIT2  uses three different sets of variables to characterize trajectories. Their definitions and the conversions between them is discussed below.

We start by introducing the conventions for labelling planes. A plane in three-space is defined by three vectors: we call its origin $\vec{ o}$ (0,0,0), and define the two orthogonal unit vectors $u$, $v$, giving the directions of the coordinate axes; the normal of the plane points in the direction of $\vec{n}$ =  $\vec{u} \times \vec{v}$.  $\vec{v}$, $\vec{u}$ and $\vec{n}$ identify a left-handed system, respectively, as shown  in Fig.~\ref{fig:detPlane},  Sec.~\ref{sec-repr}.

 The three sets of variables are:
\begin{itemize}
\item \emph{6D-coordinates}. This is the common set used for user interfacing. It contains the cartesian coordinates for the position and the momentum vector, labeled as:
\begin{center}
({\bf x, p}).
\end{center}
These, together with an independently provided charge $q$, specify a helix uniquely.
\item \emph{local coordinates}. These are the five coordinates used in track fitting. They are given relative to a plane with local orthogonal coordinates $u$ and $v$. The coordinates are labeled as:\\
\begin{center}
$(q/p, u^\prime, v^\prime, u, v)$
\end{center}

Here $q$ is the charge of the particle; $p$ is its momentum; $u^\prime$, $v^\prime$ are the direction cosines relative to the plane; $u$, $v$ are the local coordinates on the plane.

These variables are supplemented by a variable $\sigma = \pm 1$, indicating whether the vertical component of the track enters the supporting plane parallel or antiparallel to its normal vector.
Conversion to the 6D-coordinates is achieved as follows:\\
\begin{center}
\( \left( \begin{array}{c}
q/p \\
u^\prime \\
v^\prime\\
u \\
v \end{array} \right)\) 
$\rightarrow$ 
\( \left( \begin{array}{ll}
\bf{O} + u \bf{U} + v \bf{V}\\
\frac{\sigma \bf{N} + u^\prime \bf{U} + v^\prime \bf{V}}{\sqrt{1+u^{\prime 2} + v^{\prime 2}}} \times p \end{array} \right)\) 
\end{center}

\item \emph{global coordinate}. These are the seven coordinates used in track extrapolation: the position {\bf x}, the direction unit vector {\bf T} and charge over momentum of the track, written as:
\begin{center}
({\bf x}, {\bf T}, q/p)
\end{center}
Conversion to the 6D-coordinates is performed by the mapping:
\begin{center}
({\bf x}, {\bf T}, q/p) $\rightarrow$ ({\bf x}, $p${\bf T}).
\end{center}
These coordinates contain the complete helix information.
Additionally, we can represent a track by its helix parameters, which are labeled as:
\begin{center}
($d_0, \phi, \omega, z_0, tan \lambda$)
\end{center}
and which are defined following the conventions used at BaBar~\cite{dave}.
\end{itemize}
Jacobians are used for the various possible coordinate transformations. For the mapping between two sets of vectors, $x \rightarrow y$, we define the Jacobian matrix as:
\begin{equation}
J(x \rightarrow y) = (\frac{\partial y}{\partial x}),  
\end{equation}
where the $y$ labels the rows and the $x$ labels the columns. This, together with matrix multiplication, gives the usual chain rule for the mapping $x \rightarrow y \rightarrow z$,
\begin{equation}
J(x \rightarrow z) = J(y \rightarrow z)J(x \rightarrow y)
\end{equation}
and we obtain the usual error propagation formula:
\begin{equation}
C_y = J(x \rightarrow y) \cdot C_x \cdot J(x \rightarrow y)^T
\end{equation}
which expresses the covariance matrix of the $y$ in terms of the error matrix of the $x$ together with the Jacobian of the mapping between them.

\begin{equation}
J(local \rightarrow global) =
\begin{pmatrix}
0  & 0  & 0  & U_x  & V_x \\
0  & 0  & 0  & U_y  & V_y \\
0  & 0  & 0  & U_z  & V_z \\
0  & \frac{-u^\prime \sigma N_x - u^\prime v^\prime V_x + (1+v^{\prime 2})U_x}{(1 + u^{\prime 2} + v^{\prime 2})^{3/2}}   & \frac{-v^\prime \sigma N_x - u^\prime v^\prime U_x + (1+u^{\prime 2})V_x}{(1 + u^{\prime 2} + v^{\prime 2})^{3/2}}   & 0  & 0 \\
0  & \frac{-u^\prime \sigma N_y - u^\prime v^\prime V_y + (1+v^{\prime 2})U_y}{(1 + u^{\prime 2} + v^{\prime 2})^{3/2}}   & \frac{-v^\prime \sigma N_y - u^\prime v^\prime U_y + (1+u^{\prime 2})V_y}{(1 + u^{\prime 2} + v^{\prime 2})^{3/2}}   & 0  & 0 \\
0  & \frac{-u^\prime \sigma N_z - u^\prime v^\prime V_z + (1+v^{\prime 2})U_z}{(1 + u^{\prime 2} + v^{\prime 2})^{3/2}}   & \frac{-v^\prime \sigma N_z - u^\prime v^\prime U_z + (1+u^{\prime 2})V_z}{(1 + u^{\prime 2} + v^{\prime 2})^{3/2}}   & 0  & 0 \\
 \end{pmatrix}
\end{equation}
\begin{equation}
J(6D \rightarrow global) =
\begin{pmatrix}
1   & 0   & 0   & 0   & 0   & 0 \\  
0   & 1   & 0   & 0   & 0   & 0 \\ 
0   & 0   & 1   & 0   & 0   & 0 \\ 
0   & 0   & 0   & \frac{p_y^2 + p_z^2}{p^{3/2}} & -\frac{p_xp_y}{p^{3/2}}       & -\frac{p_xp_z}{p^{3/2}} \\
0   & 0   & 0   &  -\frac{p_xp_y}{p^{3/2}}     &  \frac{p_x^2 + p_z^2}{p^{3/2}} &  -\frac{p_yp_z}{p^{3/2}} \\
0   & 0   & 0   &  -\frac{p_xp_z}{p^{3/2}}     &  -\frac{p_yp_z}{p^{3/2}}       &  \frac{p_x^2 + p_y^2}{p^{3/2}} \\
0   & 0   & 0   &  -\frac{q p_x}{p^{3/2}}      &  -\frac{q p_y}{p^{3/2}}       &  -\frac{q p_z}{p^{3/2}}\end{pmatrix} 
\end{equation}
\begin{equation}
J(global \rightarrow 6D) =
\begin{pmatrix}
1   & 0   & 0   & 0   & 0   & 0   & 0\\  
0   & 1   & 0   & 0   & 0   & 0   & 0\\ 
0   & 0   & 1   & 0   & 0   & 0   & 0\\ 
0   & 0   & 0   & p   & 0   &  0  & -p^2 q T_x\\
0   & 0   & 0   & 0   & p   &  0  & -p^2 q T_y\\
0   & 0   & 0   & 0   & 0   &  p  & -p^2 q T_z \end{pmatrix} 
\end{equation}
The important special cases, where the plane is orthogonal to the track, $i.e.$ $u^\prime$ = $v^\prime$ = 0, delivers:
\begin{equation}
J(local \rightarrow 6D) = J(global \rightarrow 6D) \cdot J(local \rightarrow global) =  \\
\begin{pmatrix}
0   & 0   & 0 & U_x & V_x \\
0   & 0   & 0 & U_y & V_y \\
0   & 0   & 0 & U_z & V_z  \\
-p^2 q \sigma N_x  & U_x & V_x & 0 & 0 \\
-p^2 q \sigma N_y  & U_y & V_y & 0 & 0 \\
-p^2 q \sigma N_z  & U_z & V_z & 0 & 0 \end{pmatrix} 
\end{equation}
In order to express efficiently the Jacobian for conversion from 6D to local coordinates, we express the momentum vector in its components ($p_U$, $p_V$, $p_N$), where, $e.g.$ $p_U \equiv {\bf p} \cdot {\bf U}$, and $\sigma$ = sgn(${\bf p_N}$). Then:
\begin{equation}
J(6D \rightarrow local) = 
\begin{pmatrix}
0   & 0   & 0   & -\frac{q(p_N N_x + p_U U_x + p_V V_x)}{p^3}   & -\frac{q(p_N N_y + p_U U_y + p_V V_y}{p^3}   & -\frac{q(p_N N_z + p_U U_z + p_V V_z}{p^3} \\ 
0   & 0   & 0   & \frac{\sigma (p_N U_x - p_U N_x)}{p_N^2}   & \frac{\sigma (p_N U_y - p_U N_y}{p_N^2}   &  \frac{p_N U_z - p_U N_z}{p_N^2} \\ 
0   & 0   & 0   & \frac{\sigma (p_N V_x - p_v N_x}{p_N^2}   & \frac{\sigma (p_N V_y - p_V N_y}{p_N^2}   &  \frac{p_N V_z - p_V N_z}{p_N^2} \\ 
U_x & U_y  & U_z &          0  & 0           &  0         \\
V_x & V_y  & V_z &          0  & 0           &  0      \end{pmatrix}   
\end{equation}
\begin{equation}
J(global \rightarrow local) = 
\begin{pmatrix}
0   & 0   & 0   & 0    & 0    & 0     & 1 \\
0   & 0   & 0   & \frac{({\bf T \cdot N})U_x -({\bf T \cdot N})N_x}{({\bf T \cdot N})^2}\sigma & \frac{({\bf T \cdot N})U_y - ({\bf T \cdot N})N_y}{({\bf T \cdot N})^2}\sigma & \frac{({\bf T \cdot N})U_z - ({\bf T \cdot N})N_z}{({\bf T \cdot N})^2}\sigma  & 0 \\
0   & 0   & 0   &\frac{({\bf T \cdot N})V_x -({\bf T \cdot N})N_x}{({\bf T \cdot N})^2}\sigma & \frac{({\bf T \cdot N})V_y - ({\bf T \cdot N})N_y}{({\bf T \cdot N})^2}\sigma & \frac{({\bf T \cdot N})V_z - ({\bf T \cdot N})N_z}{({\bf T \cdot N})^2}\sigma  & 0 \\
U_x & U_y  & U_z &          0  & 0           &  0    & 0     \\
V_x & V_y  & V_z &          0  & 0           &  0    & 0  \end{pmatrix}   
\end{equation}
Again, the special case where momentum is orthogonal to the plane in question ({\bf T $\cdot$ N} = $\frac{1}{p}$p$\cdot$N = $\sigma$ = $\pm$1), gives:
\begin{equation}
J(6D \rightarrow local) = 
\begin{pmatrix}
0   & 0   & 0   & -\frac{q}{p^2}N_x   &  -\frac{q}{p^2}N_y  &  -\frac{q}{p^2}N_z\\ 
0   & 0   & 0   & \sigma \frac{U_x}{p}   &   \sigma \frac{U_y}{p}  &  \sigma \frac{U_z}{p}  \\ 
0   & 0   & 0   &  \sigma \frac{V_x}{p}  &  \sigma \frac{V_y}{p}  &  \sigma \frac{V_z}{p} \\ 
U_x & U_y  & U_z &          0  & 0           &  0         \\
V_x & V_y  & V_z &          0  & 0           &  0      \end{pmatrix}
\end{equation}
and:
\begin{equation}
J(global \rightarrow local) = 
\begin{pmatrix}
0   & 0   & 0   &  0  &  0  & 0 & 1 \\ 
0   & 0   & 0   & \sigma U_x & \sigma U_y   & \sigma U_z & 0 \\ 
0   & 0   & 0   & \sigma V_x & \sigma V_y   & \sigma V_z & 0  \\ 
U_x & U_y  & U_z &         0  & 0         0 &  0         & 0 \\
V_x & V_y  & V_z &         0  & 0         0 &  0         & 0 \end{pmatrix}
\end{equation}

As consistency check, one can try to calculate the transformation: $local \rightarrow global \rightarrow local$: the Jacobian of this transformation is indeed equal to 0; the converse transformation, $e.g.$  $global \rightarrow local \rightarrow global$, has singular Jacobian. Its null space is spanned by: ($0, 0, 0, {\bf T_x}, {\bf T_y}, {\bf T_z}$), (${\bf N_x}, {\bf N_y}, {\bf N_z}, 0, 0, 0, 0$), which comes about because T is normalized (this brings about the first direction of the null space) and because a point in local coordinates is constrained to a plane (the second direction). 
\section{Averaging}
\label{app:averaging}

Given two measurements of the same $n$-dimensional quantity and their
respective covariances, $(x_1,C_1)$ and $(x_2,C_2)$, the two
measurements can be combined to a least-likelihood estimate $\tilde x$
by minimizing the $\chi^2$ of the combination, $i.e.$
\begin{equation}
  \label{eq:5}
  \chi^2 = (\tilde x-x_1)^TC_1^{-1}(\tilde x-x_1) +(\tilde x-x_2)^TC_2^{-1}(\tilde x-x_2) \equiv \min_{\tilde x}.
\end{equation}
The solution is the well-known formula:
\begin{equation}
  \label{eq:6}
  \tilde x = (C_1^{-1}+C_2^{-1})^{-1}(C_1^{-1}x_1+C_2^{-1}x_2),
\end{equation}
and the covariance of $\tilde x$ is by linear error propagation:
\begin{equation}
  \label{eq:7}
  \tilde C = (C_1^{-1}+C_2^{-1})^{-1}.
\end{equation}
We can simplify the evaluation of these textbook formulas
significantly in the following manner, applying a technique from
Ref.~\cite{golub:1965}: covariance matrices are positive definite
matrices.  As such, they can be written $C=S^TS$ where $S$ is some
matrix which can be chosen to be upper triangular.  In the case where
$C$ is diagonal, the entries of $C$ correspond to the squared errors
$\sigma^2$ and the entries of $S$ correspond to the errors $\sigma$.
It is in this sense that $S$ is referred to as a square root of $C$.
It is commonly evaluated by Cholesky decomposition.  Since $S$ is a
upper triangular matrix its inverse, and thus the inverse of $C$, can
be evaluated directly by means of forward substitution.  Likewise
products of the form $S^{-1}x$ can be evaluated by forward
substitution without first evaluating the inverse.  Our goal is to
replace the matrices in Eq.~\eqref{eq:6} and Eq.\eqref{eq:7} by their square
roots and use those directly, thus operating with the errors instead
of the squared errors.

Starting with the inner part of \eqref{eq:7}, we can write:
\begin{align}
  \label{eq:9}
  C_1^{-1}+C_2^{-1} & =
  {S_1^{-1}}^TS_1^{-1}+{S_2^{-1}}^TS_2^{-1}\\
  & = ({S_1^{-1}}^T,{S_2^{-1}}^T)\left(\begin{array}{c}{S_1^{-1}}\\
                                       {S_2^{-1}}\end{array}\right),
\end{align}
where on the second line we use block matrix notation, $i.e.$ the
matrix on the right is $2n\times n$.  As such, there is an $2n\times
2n$-dimensional orthogonal matrix $Q$, $QQ^T=1$, which converts it to
an upper triagonal matrix, written explicitly:
\begin{equation}
  \label{eq:12}
  \left(\begin{array}{c}{S_1^{-1}}\\
        {S_2^{-1}}\end{array}\right) = Q\left(\begin{array}{c}R\\
        0\end{array}\right),
\end{equation}
where $R$ is $n\times n$.  This operation is the
$QR$ decomposition of the left-hand side, and we can write:
\begin{align}
  \label{eq:10}
  C_1^{-1}+C_2^{-1} & = (R^T,0)QQ^T\left(\begin{array}{c}R\\
                                           0\end{array}\right)\\
& = R^T R.
\end{align}
Going back to Eq. \eqref{eq:7}, we have:
\begin{equation}
  \label{eq:11}
  \tilde C = R^{-1} {R^{-1}}^T.
\end{equation}
In other words, we can obtain the square root of $\tilde C$ directly
from the square roots of $C_1$, $C_2$ by stacking them and performing
a $QR$ decomposition.  This increases the numerical range in the
presence of vastly different errors, as we no longer operate on the
squared errors but on the errors themselves.  Additionally, any matrix
of the form in Eq.~\eqref{eq:11} is always invertible and positive
definite which prevents catastrophic conditions in subsequent
numerical operations.

It is perhaps revealing to consider the similarity to the Pythagorean
theorem which gives the length of a vector $(a,b)^T$ as $a^2+b^2=c^2$.
The corresponding orthogonal transformations in this case are the
rotations which take the vector $(a,b)^T$ to $(\pm
\sqrt{a^2+b^2},0)^T$.  The computational gain comes from the fact that
the orthogonal operation disappears because the relevant quantities
are squares which always lead to factors $QQ^T$ or $Q^TQ$ (for the
inverse matrices).

In order to evaluate $\tilde x$, we insert those into Eq. \eqref{eq:6}, and
obtain:
\begin{align}
  \label{eq:13}
  \tilde x & = R^{-1}{R^{-1}}^T
               (R^T,0)Q\left(\begin{array}{c}S_1^{-1}x_1\\
                               S_2^{-1}x_2\end{array}\right) \\
 & = R^{-1} \bar Q \left(\begin{array}{c}S_1^{-1}x_1\\
                               S_2^{-1}x_2\end{array}\right),
\end{align}
where $\bar Q$ refers to the upper $n$ rows of $Q$.  It never
explicitly appears in the implementation, instead the right-hand
multiplication is evaluated simultaneously with the $QR$ decomposition
in Eq.~\eqref{eq:12}, using the algorithm of Ref.~\cite{Businger:1965}.

In practice this algorithm is not only preferable because of its
numerical properties, we also measured it to perform slightly faster
than a naive implementation of Eqs.~\eqref{eq:6} and~\eqref{eq:7}.

\section{Proof of Square-root Kalman filter formula}
\label{sec:proof-square-root}

In a similar vein to Appendix~\ref{app:averaging}, we want to express the
expressions for $\bfp_{i|i}$ and the square root of $C_{i|i}=\tilde
S^T\tilde S$ in Eq.\eqref{eq:15} in terms of the $n\times n$ square roots
of $C_{i|i-1}=S^T S$ and the $m\times m$ measurement covariance
$V_i=R^TR$.
Following Ref.~\cite{Anderson:2005}, we want to show that
we can find corresponding expressions by $QR$ decomposition of the
matrix:
\begin{equation}
  \label{eq:16}
  \left(\begin{array}{cc} R & 0 \\ SH^T & S\end{array}\right)
= Q\left(\begin{array}{cc}U & X \\ 0 & W\end{array}\right),
\end{equation}
where $QQ^T=1$, $U$ and $W$ are upper triagonal matrices, and the
block sizes on the right-hand and left-hand sides are equal.  For sake
of brevity we write $H=H_i$, $V=V_i$, $C=C_{i|i-1}$.

We want to show that $W=\tilde S$ and
$X^T{U^{-1}}^T=CH^T(V+HCH^T)^{-1}$.  In order
to do this, we square both sides of \eqref{eq:16}, $i.e.$ multiply each side
with its transposed from the left.  This gives:
\begin{equation}
  \label{eq:17}
  \left(\begin{array}{cc}V+HCH^T & HC\\
          CH^T & C\end{array}\right) =
      \left(\begin{array}{cc}U^TU & U^TX\\
              X^TU & X^TX+W^TW\end{array}\right).
\end{equation}
From this we can read off $U^TU=V+HCH^T$, $U^TX=HC$.  We can now find
$X^TX$ by strategical insertion of the $m\times m$ identity matrix
followed by a rearrangement of parentheses,
\begin{equation}
  \label{eq:18}
  X^TX=X^T\cdot 1\cdot X = X^T (U U^{-1}) ({U^{-1}}^T U^T) X = (X^T
  U)(U^TU)^{-1}(U^TX).
\end{equation}
By identifying the corresponding entries in~\eqref{eq:17}, we see that
\begin{equation}
  \label{eq:19}
  X^TX = CH^T(V+HCH^T)^{-1} HC.
\end{equation}
Inserting this into the lower right block of~\eqref{eq:17} and
comparing to~\eqref{eq:15}, we find that:
\begin{equation}
  \label{eq:21}
  W^TW=C-CH^T(V+HCH^T)^{-1}HC=\tilde S^T\tilde S
\end{equation}
and thus identify $W=\tilde S$, the desired square root of the updated
covariance matrix.  Finally, by another strategic insertion of 1 we
find:
\begin{equation}
  \label{eq:20}
  X^T{U^{-1}}^Tr_i=X^TU(U^T U)^{-1}r_i=CH^T(V+HCH^T)^{-1}r_i,
\end{equation}
$i.e.$ the update of the state.  Observe how the more familiar formulas
are obtained by inserting additional, cancelling terms in
Eqs.~\eqref{eq:18} and\ \eqref{eq:20}.  These cancellations are what
degrade the numerical behavior of the more commonly used formulas.

\section{List of acronyms}
\label{acronym}
BASF = Belle Analysis Software Framework\\
CDC  = Central Drift Chamber\\
CKF  = Combined Kalman Filter\\
DAF  = Deterministic Annealing Filter\\
EM =  Electro-Magnetic calorimeter\\
FAIR = Facility for Antiproton and Ion Research\\
GENFIT = General Fitter\\
GSI  = Gesellschaft f\"ur Schwerionenforschung (Helmholtz Centre for Heavy Ion Research)\\
GBL = Global Broken Line\\
POCA = Point of Closest Approach\\
PXD = Pixel Detector\\
RAVE = Reconstruction  of  vertices  in  Abstract Versatile Environments\\
SVD = Silicon Vertex Detector\\
STT = Straw-Tube Tracker\\
TOP = Time-Of-Propagation detector\\

%
%
\end{document}